\documentclass[multphys,vecphys]{svmult}
\usepackage{makeidx}         % allows index generation
\usepackage{graphicx}        % standard LaTeX graphics tool
\usepackage{multicol}        % used for the two-column index
\usepackage[bottom]{footmisc}% places footnotes at page bottom
\usepackage{bm}

\makeindex             % used for the subject index
\begin{document}

\title*{Quantum Gates and Decoherence}
\author{Stefan Scheel\inst{1}
\and Jiannis Pachos\inst{2}
\and E.A. Hinds\inst{1}
\and Peter L. Knight\inst{1}}
\institute{Quantum Optics and Laser Science, Blackett Laboratory,
Imperial College London, Prince Consort Road, London SW7 2BW, United
Kingdom 
\texttt{s.scheel@imperial.ac.uk}
\and DAMTP, Centre for Mathematical Sciences, Wilberforce Road,
Cambridge CB3 0WA, United Kingdom.
}
\maketitle

%%%%%%%%%%%%%%%%%%%%%%%%%%%%%%%%%%%%%%%%%%%%%%%%%%%%%%%%%%%%%%%%%%%%%%
\begin{quote}
"God forbid that we should give out a dream of our own imagination for
a Pattern of the World."\\ \hfill --- Francis Bacon, Novum Organum
\end{quote}

%%%%%%%%%%%%%%%%%%%%%%%%%%%%%%%%%%%%%%%%%%%%%%%%%%%%%%%%%%%%%%%%%%%%%%
\section{Introduction}
\label{sec:intro}

\begin{figure}[ht]
\centerline{\includegraphics[width=7cm]{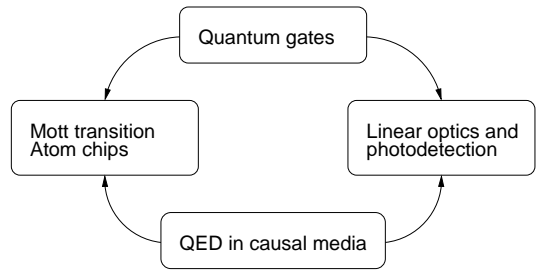}}
\caption{\label{fig:introfig} Connections between the subject areas
covered.}
\end{figure}

In this article we will be concerned with some possible physical
realizations of quantum gates that are useful for quantum information
processing. After a brief introduction into the subject, in
Sec.~\ref{sec:atoms} we will focus on a particular way of using atoms
in one-dimensional optical lattices as carriers of the quantum
information. In Sec.~\ref{sec:photons} on the contrary, the
information carriers will be photons that interact via effective
nonlinearities which arises from mixing at passive linear optical
elements and postselection through photodetection. These two seemingly
different implementations have in common that their decoherence
mechanism is described by a single theory, namely that of quantum
electrodynamics in causal media which will be the subject of
Sec.~\ref{sec:media}. Figure~\ref{fig:introfig} should serve as an
overview of the subject areas covered.
%

%%%%%%%%%%%%%%%%%%%%%%%%%%%%%%%%%%%%%%%%%%%%%%%%%%%%%%%%%%%%%%%%%%%%%%
\subsection{Why quantum information processing?}

Apart from the academically driven curiosity to learn more about
information in a quantum-mechanical setting, there is very urgent
practical need to investigate information processing and computing
from a quantum theory point of view. Over the last 30 years the number
of transistors on an integrated circuit, i.e. the complexity of the
computer made up from those chips, doubles roughly every 18
months. This empirical behaviour is famously known as Moore's first
law (after the co-founder of Intel Corp., G.~Moore).
\index{Moore's law}
An extrapolation
reveals that by the year 2017 a bit of information will need to be
encoded in a single atom. Even if Moore's law breaks down before we
eventually come to this point, at the current growth rate, by the year
2012 the dimensions of logical elements will be so small that quantum
effects upon computation cannot be neglected any longer.
\footnote{The most recent Intel Pentium 4 processor already contains
features of the size of 90nm, i.e. 900\AA. The lithography used to
generate such structures therefore uses XUV light!} 
Thus, technological progress necessitates the study of what
implications quantum theory has on computation.

Another reason for looking deeper into quantum computing lies in the
potential ability to simulate the temporal evolution of (possibly
chaotic) quantum systems. The intrinsic and potentially massive
parallelism of quantum computers that stems from linearity of quantum
mechanics and the resulting superposition principle, would allow to
investigate quantum Hamiltonian systems without the need for an
exponential temporal overhead on a classical computer (see
e.g. \cite{simulation}). 

%%%%%%%%%%%%%%%%%%%%%%%%%%%%%%%%%%%%%%%%%%%%%%%%%%%%%%%%%%%%%%%%%%%%%%
\subsection{Quantum gates vs. classical gates}

\begin{quote}
"Information is physical." \\ \hfill --- Rolf Landauer
\end{quote}

The most important difference between a logical gate known from
classical information processing and a quantum gate is its
reversibility or conservation of information. In order to see what
that means, let us consider for example the logical NAND gate whose
action of logical 0's and 1's are given by the truth table
\ref{tab:nand}. \index{classical NAND gate}
\begin{table}
\centering
\caption{Truth table of the logical NAND gate \label{tab:nand}}
\begin{tabular}{c|c}
\hline\noalign{\smallskip}
inputs & output  \\
\noalign{\smallskip}\hline\noalign{\smallskip}
0\,0&1\\0\,1&1\\1\,0&1\\1\,1&0\\
\noalign{\smallskip}\hline
\end{tabular}
\end{table}
Equivalently, we can describe it by the action of
$\overline{X_1\wedge X_2}$ on two Boolean variables $X_1$ and $X_2$.
This gate is generic for classical gates in that it has two (or more)
inputs and only a single output. In quantum mechanics, we are allowed
to form linear combinations or superpositions of the logical basis
states,
i.e. $c_0|00\rangle+c_1|01\rangle+c_2|10\rangle+c_3|11\rangle$. The
formal application of the classical NAND on this quantum superposition
would result in a state $\propto
c_3|0\rangle+(c_0+c_1+c_2)|1\rangle$.
The resulting state contains far less information than the
original state. Whereas before the gate operation all the weight
coefficients $c_i$ gave us information about the quantum state, after
the action of the classical NAND only the \textit{sum} $c_0+c_1+c_2$
plays a r\^{o}le in determining the weight of the logical
$|1\rangle$. All the information about the mutual \textit{differences}
are lost during the operation.
However, this state is obviously not properly normalized anymore. The
formal action of a classical gate is therefore actually an ill-posed
operation in quantum mechanics. In fact, there is no way of making
sense of classical logical operations acting on quantum superpositions. 

In quantum information processing, all operations have to be
unitary. This means on one hand that the number of output degrees of
freedom must equal the number of input degrees of freedom, and on the
other hand it means that no information is lost about an initial
superposition. In general, we could allow even more general operations
such as completely positive  maps of which the unitary operations are
a subset. However, not even using the correct number of inputs and
outputs together with no loss of information is sufficient to define a
quantum operation. For example, let us consider an operation in which
a given quantum state $c_0|0\rangle+c_1|1\rangle$ is transformed into
its orthogonal complement, hence into the state
$-c_1^\ast|0\rangle+c_0^\ast|1\rangle$. This operation could be
thought of as being the analogue of the classical NOT operation which
is defined as $X\mapsto\overline{X}:\overline{X}\wedge X=0$. However,
it turns out that the quantum-NOT operation is not a completely
positive map and therefore cannot be implemented by any quantum
circuit. In fact, the map describing this transformation is
anti-unitary \cite{Buzek99} with determinant $-1$.
\footnote{In this context, the notion of universal gate has been
invented. A universal NOT-gate would be the one that, when averaging
over all input state,  comes closest to the NOT-operation.}

\paragraph{Universal set of quantum gates}

Now we have shown that there is no strict connection between classical
logical gates and quantum mechanics, we will next briefly discuss which
operations are compatible with quantum mechanics. Let us first look
into operations acting on a single logical qubit state
$c_0|0\rangle+c_1|1\rangle$. The dynamical group associated with this
state is the unitary group SU(2) whose action can be given in terms of
its generators, the Pauli spin operators $\hat{\sigma}_i$ ($i=x,y,z$),
as $\hat{U}=e^{i\bm{\alpha}\cdot\hat{\bm{\sigma}}}$. Here,
$\bm{\alpha}$ denotes the direction of the rotation axis and the
associated rotation angle. By decomposing $\hat{U}$ into exponential
factors, one can derive the so-called Euler decomposition
$\hat{U}=e^{ia_1\hat{\sigma}_z}e^{ia_2\hat{\sigma}_y}e^{ia_3\hat{\sigma}_z}$
\index{Euler decomposition of SU(2)}
which just describes a sequence of elementary rotations. The Pauli
operators have the following representation in the single-qubit basis:
\begin{eqnarray}
\hat{\sigma}_x &=& |0\rangle\langle 1|+|1\rangle\langle 0| \,,\\
\hat{\sigma}_y &=& i(|0\rangle\langle 1|-|1\rangle\langle 0|) \,,\\
\hat{\sigma}_z &=& |0\rangle\langle 0|-|1\rangle\langle 1| \,.
\end{eqnarray}
\index{Hadamard gate}
An equivalent set of operators would be
$\hat{\sigma}_\pm=\hat{\sigma}_x\pm i\hat{\sigma}_y$ and
$\hat{\sigma}_z$ which results in an operator decomposition of the form
$\hat{U}=e^{ib_1(1+\hat{\sigma}_z)}e^{ib_2\hat{\sigma}_+}$
$ie^{b_3\hat{\sigma}_-}ie^{b_4(1-\hat{\sigma}_z)}$.
This result will be used later in Sec.~\ref{sec:photons} to describe
beam splitters.

By definition, the Pauli operators are enough to represent all other
single-qubit operations. An important example is the Hadamard gate
$\hat{H}$ (not to be confused with the Hamiltonian), 
\begin{equation}
\hat{H} = \frac{1}{\sqrt{2}} \Big[
\left( |0\rangle +|1\rangle \right) \langle 0| +
\left( |0\rangle -|1\rangle \right) \langle 1| \Big] \,,
\end{equation}
which transform logical states into coherent superpositions. From the
definition of the Pauli operators it is obvious that the Hadamard gate
can be written as $\hat{H}=(\hat{\sigma}_x+\hat{\sigma}_z)/\sqrt{2}$.

\begin{figure}[ht]
\centerline{\includegraphics[width=6cm,clip=]{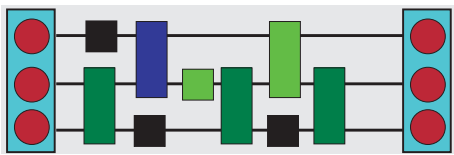}}
\caption{\label{fig:cat} A quantum network can be build from a
universal set of quantum gates. Here we sketch a three bit register
which is manipulated by a sequence of one and two qubit gates to
realize a chosen unitary transformation which effects the relevant
algorithm.}
\end{figure}

As for two-qubit gates, the structure of the unitary operator
associated with the dynamical symmetry is not so obvious. We will
therefore restrict our attention to some particularly useful
gates. One of the most important nontrivial two-qubit gates is the
controlled-phase gate $\hat{C}_\varphi$ whose truth table is given in
Table~\ref{tab:cphase}.
\index{controlled-phase gate}
\begin{table}
\centering
\caption{Truth table of the controlled-phase gate \label{tab:cphase}}
\begin{tabular}{lll}
\hline\noalign{\smallskip}
input & output  \\
\noalign{\smallskip}\hline\noalign{\smallskip}
$|00\rangle$ & $|00\rangle$ \\
$|01\rangle$ & $|01\rangle$ \\
$|10\rangle$ & $|10\rangle$ \\
$|11\rangle$ & $e^{i\varphi}|11\rangle$ \\
\noalign{\smallskip}\hline
\end{tabular}
\end{table}
The controlled-phase gate (or its special case the
controlled-$\hat{\sigma}_z$ gate with $\varphi=\pi$), together with
the set of Pauli operators, forms a universal set of quantum gates
\cite{DiVincenzo95} by which we mean that all possible quantum
networks can be build up from them. Therefore, if one is able to build
these four gates, one can generate arbitrary quantum operations by
concatenating them (Fig.~\ref{fig:cat}).

For example, another nontrivial two-qubit gate is the controlled-NOT
(or CNOT) under whose action the logical states $|10\rangle$ and
$|11\rangle$ are exchanged and the other left untouched.
\index{controlled-NOT gate}
\footnote{The associated operator can be written as
$|0\rangle\langle 0|\otimes\hat{I}+|1\rangle\langle 1|\otimes\hat{\sigma}_z$.}
This operator can be realized by the combination
$\hat{H}_1\hat{C}_\pi\hat{H}_1$, i.e. by two Hadamard gates acting on
mode 1 and a controlled-$\hat{\sigma}_z$ gate. Other two-qubit gates
of importance are the swap in which the qubits are simply exchanged
and the square root of swap which serves as an entangling gate.

%%%%%%%%%%%%%%%%%%%%%%%%%%%%%%%%%%%%%%%%%%%%%%%%%%%%%%%%%%%%%%%%%%%%%%
\section{Atomic realisation --- atom chips and the Mott transition in
optical lattices}
\label{sec:atoms}

There are many possible realizations of quantum bits: laser-cooled
trapped ions, cold trapped neutral atoms, atoms in high-Q single-mode
cavities as well as condensed matter candidates such as quantum dots
or Josephson junctions. For any one of these to be viable, the effects
of the environment (dissipation and temperature) must be minimized, if
not entirely eradicated. This simply means we need to use cold state
initialization.

In this section, we will describe a possible way of implementing
quantum computation with cold atom technology. This includes the
application of optical lattices in a sufficiently cold cloud of atoms
showing Bose-Einstein condensation (BEC). The lasers which generate
the periodic spatially varying trapping potential through the AC Stark
effect create an optical lattice strong enough to induce a quantum
phase transition from the superfluid phase that characterises the  BEC
to the Mott insulator phase, characterised by a regular structure of
one atom per lattice site that can serve as a quantum register. A
universal set of quantum gates can then be realised by manipulations
of the lattice potential with additional laser fields.

%%%%%%%%%%%%%%%%%%%%%%%%%%%%%%%%%%%%%%%%%%%%%%%%%%%%%%%%%%%%%%%%%%%%%%
\subsection{Bose--Einstein condensates and the Mott transition}

In recent years it has been realized that Bose--Einstein
condensates (BECs for short)
\index{Bose--Einstein condensate}
can undergo a phase transition if loaded
into a three-dimensional periodic potential which for example can be
realized by standing-wave optical fields \cite{Jaksch98}. 
That is, one starts off with a BEC in its superfluid phase in which
the relative phases (or rather correlations) between the atoms are
well-defined such that the whole ensemble of atoms can be described by
a single macroscopic wave function (in first approximation).
\begin{figure}[ht]
\centerline{\includegraphics[width=8cm]{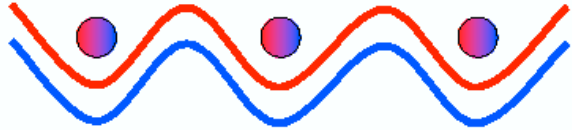}}
\caption{\label{fig:lattice} Counter-propagating laser beams induce
a periodic spatially varying trapping potential through the AC Stark shift.} 
\end{figure}
By loading this condensate into the optical lattice (see
Fig.~\ref{fig:lattice}) the 
number of atoms per lattice site is undetermined and can vary
widely. The ground-state wave function for $N$ atoms in a lattice with
$M$ sites is therefore
\begin{equation}
|\Psi_S\rangle \propto \left( \sum\limits_{i=1}^M \hat{a}_i^\dagger
\right)^N |0\rangle 
\end{equation}
where the $\hat{a}_i^\dagger$ denote creation operators of an atom at
the lattice site $i$.
However, when increasing the strength of the potential by
increasing the power of the laser beams that create the standing-wave
potential, eventually there will be a phase-transition to a state of
the condensate in which each lattice site is occupied by a fixed and
well-defined number of atoms (ideally we would like to have exactly
one atom per site). In this so-called Mott-insulator phase
\index{Mott-insulator phase}
the relative phases (or correlations) between neighboring lattice
sites are undetermined.
The ground-state wave function here looks essentially like
\begin{equation}
|\Psi_M\rangle \propto \prod\limits_{i=1}^M (\hat{a}_i^\dagger)^n
|0\rangle 
\end{equation}
where $n$ is the number of atoms per lattice site.
Experimental evidence of this phase-transition has
been obtained in the beautiful experiments described in
\cite{Greiner,Orzel}. Although a Bose--Einstein condensate really
exists only in three dimensions (since only there we find a phase
transition from a thermal cloud to a condensate), there are analogous
systems such as the quasi-condensate \cite{Petrov00} and the
Tonks--Girardeau gas \cite{Dunjko01} in one dimension that have
similar properties.

The effective interaction Hamiltonian that can be derived from the
Gross--Pitaevskii equation under the assumption that the atoms have
localized single-particle wave functions can be written as
\begin{equation}
\label{eq:bosehubbard}
\hat{H} = \frac{U}{2}\sum\limits_i \hat{n}_i(\hat{n}_i-1)
-J \sum\limits_i ( \hat{a}_i^\dagger \hat{a}_{i+1}
+\hat{a}_{i+1}^\dagger \hat{a}_i) \,.
\end{equation}
This Hamiltonian is also known in the literature as the Bose--Hubbard
Hamiltonian. \index{Bose--Hubbard Hamiltonian}
The first term with coupling strength $U$ is the collisional energy of
atoms occupying the same lattice site. Obviously, if there is zero or
just one atom per site, this term vanishes identically. The second
term is the so-called hopping term which essentially is given by the
overlap of the localized wavefunctions at neighboring sites and
describes tunneling between adjacent lattice sites with tunneling
strength $J$. If, for example, the optical lattice is formed by the
standing wave of a one-dimensional cavity mode having width $L$, the
trapping potential is given by
\begin{equation}
V(x) = -V_0 \sin^2 kx \exp\left(-\frac{2r^2}{L^2}\right)
\end{equation}
where $r$ denotes the transverse distance from the lattice axis and
$k$ the wave number of the standing wave. With this potential, the
collisional strength can be approximated by
\begin{equation}
\label{eq:collisionstrength}
U \approx \frac{4a_sV_0^{3/4}E_R^{1/4}}{\sqrt{\lambda L}}
\end{equation}
where $a_s$ is the scattering length of the atomic collisions and
$E_R$ the atomic recoil energy. An analogous derivation shows that the
tunneling rate can be expressed as
\begin{equation}
J \approx \frac{E_R}{2} \exp \left(
-\frac{\pi^2}{4}\sqrt{\frac{V_0}{E_R}}  \right) \left[
\sqrt{\frac{V_0}{E_R}} +\left( \sqrt{\frac{V_0}{E_R}} \right)^3
\right] \,.
\end{equation}

\begin{figure}[ht]
\centerline{\includegraphics[width=8cm]{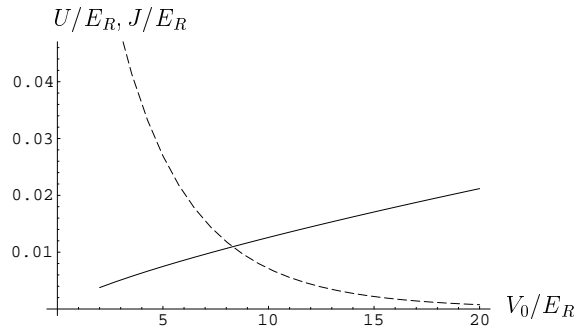}}
\caption{\label{fig:coupling} The values of $U/E_R$ (solid line) and
$J/E_R$ (dashed line) are plotted against $V_0/E_R$ for typical values
of the relevant length scales appearing in
Eq.~(\ref{eq:collisionstrength}) ($a_s=5.6$nm for $^{87}$Rb,
$L=\lambda=10\mu$m).}
\end{figure}

It is now apparent that by changing the potential depth $V_0$ one can
tune the system of atoms into either of the two (superfluid or
Mott-insulator) phases as seen in Fig.~\ref{fig:coupling}. 
For example, if $V_0$ is relatively small
compared to the recoil energy $E_R$, the tunneling rate $J$ will be
large and the atoms will be delocalized and form a
superfluid. Increasing $V_0$ means exponentially decreasing the
tunneling rate and the atoms will become stuck in their respective
potential wells and form the Mott-insulator phase.
\begin{figure}[ht]
\centerline{\includegraphics[width=6cm]{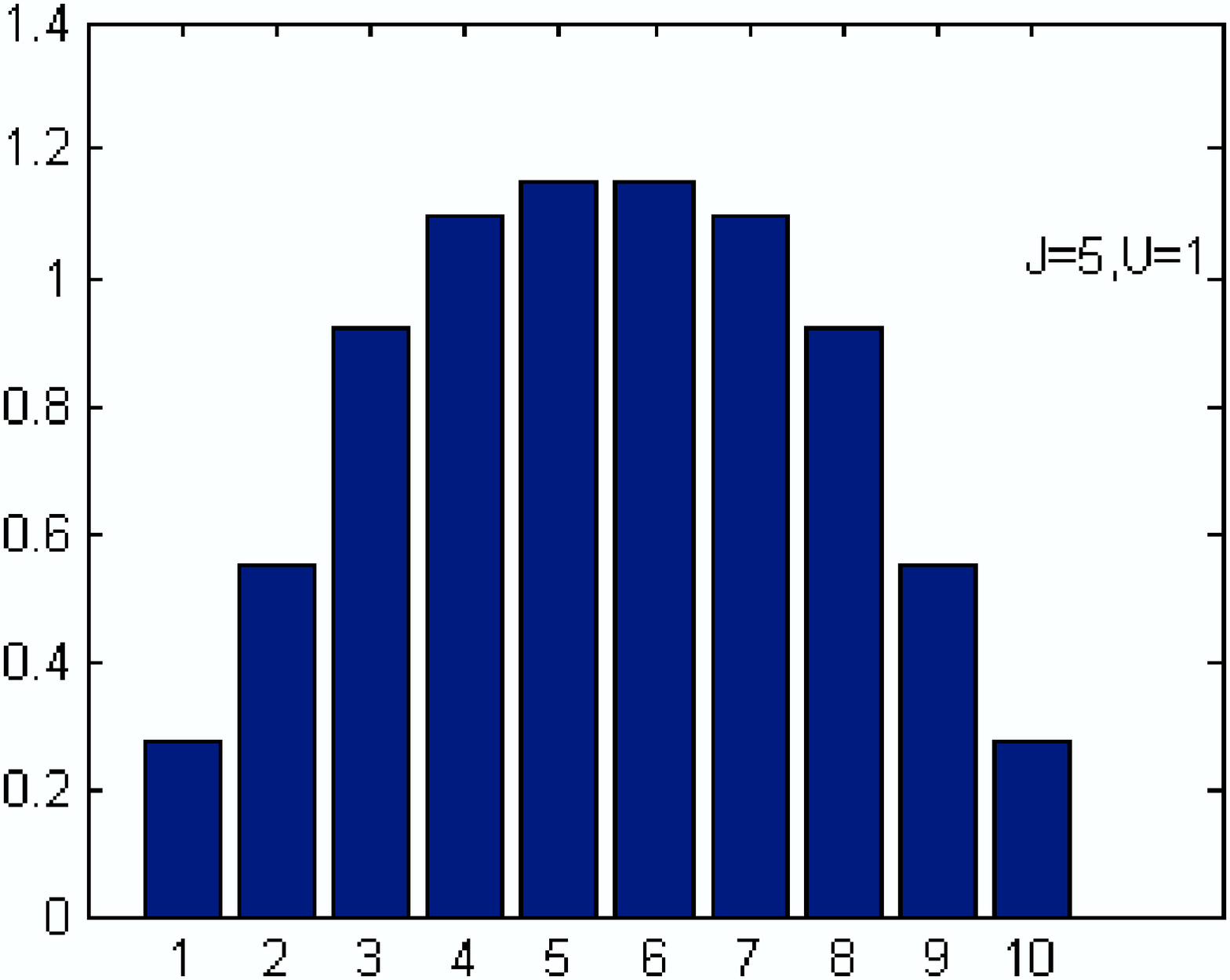}
\hfill
\includegraphics[width=6cm]{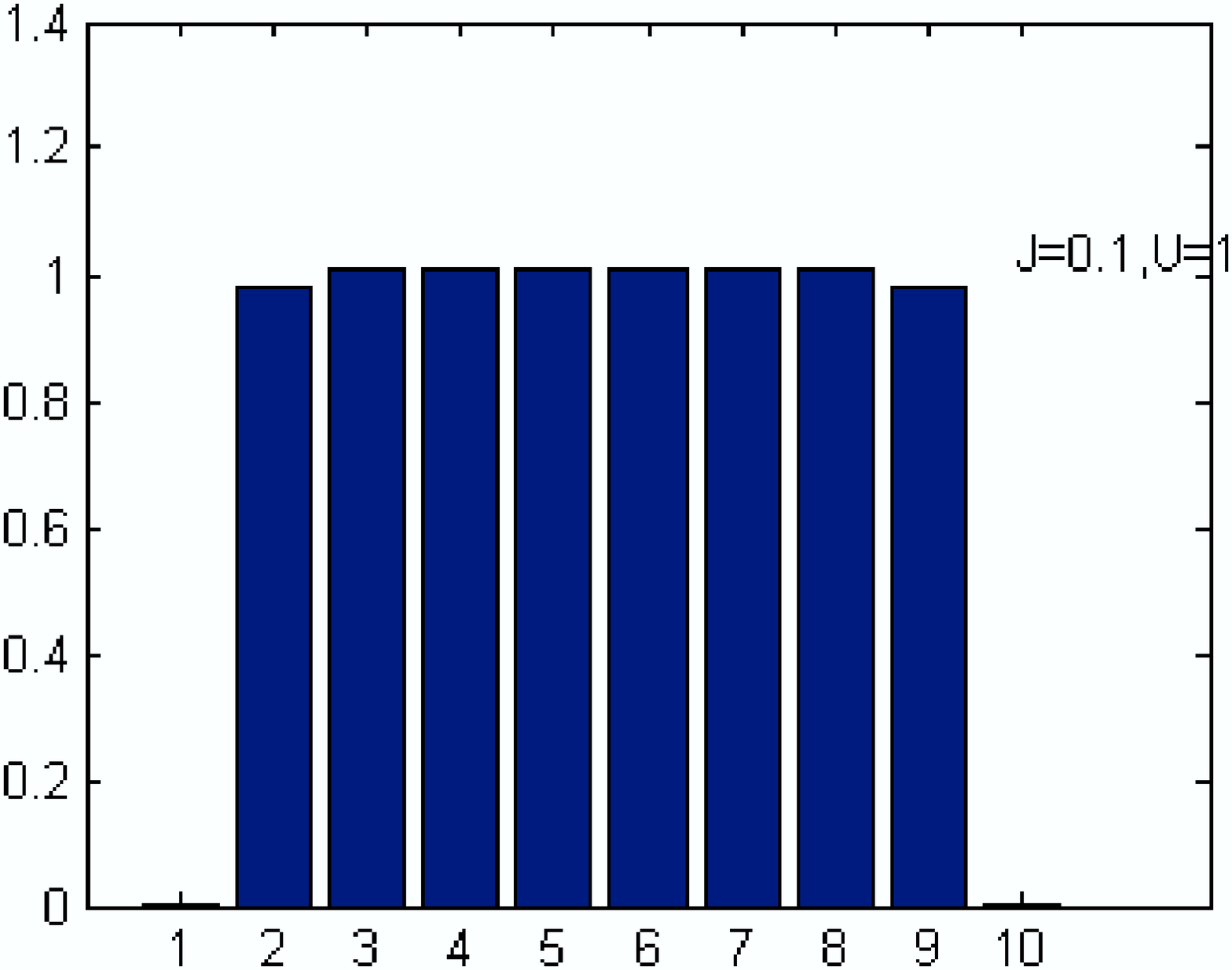}
}
\caption{\label{fig:mott} Distribution of 8 atoms among 10 lattice
sites in the limits $U/J\ll 1$ (left figure) and $U/J\gg 1$ (right
figure). }
\end{figure}
In Fig.~\ref{fig:mott} we show an example calculation for the
population distribution in a one-dimensional lattice with 10 sites
occupied by 8 atoms. The figure on the left depicts a situation in
which the collisional energy $U$ is small compared to the tunneling
rate $J$ ($U/J\ll 1$) and hence a superfluid phase exists, witnessed
by a Gaussian-like distribution. On the other hand, the figure on the
right shows the one-by-one distribution of atoms for $U/J\gg 1$ which
closely resembles a Fock state at each site. In fact, the critical
value for obtaining the phase-transition is $U/J\approx 11.6$
\cite{Fisher89}.
Note that the calculation of the ground-state wave function can only
be done numerically which amounts to computing the eigenvector
corresponding to the smallest eigenvalue of a sparse matrix of
dimension ${A+W-1 \choose A}$, where $A$ is the number of atoms in
$W$ lattice sites.

%%%%%%%%%%%%%%%%%%%%%%%%%%%%%%%%%%%%%%%%%%%%%%%%%%%%%%%%%%%%%%%%%%%%%%
\subsection{Quantum computation with a 1D optical lattice}

In order to use the Mott-insulator phase for building a quantum register,
it is advantageous to use atoms with two degenerate ground states
$|g_a\rangle$ and $|g_b\rangle$ which are coupled to each other with a
Raman transition via an excited state \cite{Pachos03}.
\index{Raman transition}
In order to trap both `species' of atoms simultaneously, an optical
lattice formed of two counterpropagating laser beams with parallel
linear polarization vectors is needed. The result of this
configuration is that the atoms are trapped in two overlapping optical
lattices with polarizations $\sigma_+$ and $\sigma_-$ each of which
can trap one of the two atomic ground states.

If we denote the creation operators of atoms in the states
$|g_a\rangle$ and $|g_b\rangle$ by $\hat{a}^\dagger$ and
$\hat{b}^\dagger$, respectively, we can write the interaction
Hamiltonian (\ref{eq:bosehubbard}) as
\begin{eqnarray}
\label{eq:bosehubbard2}
\hat{H} &=& \sum\limits_i \left[
\frac{U_{aa}}{2} \hat{n}_i^a (\hat{n}_i^a-1) 
+U_{ab} \hat{n}_i^a \hat{n}_i^b
+\frac{U_{bb}}{2} \hat{n}_i^b (\hat{n}_i^b-1) \right]
\nonumber \\ &&
-\sum\limits_i \left( J_i^a \hat{a}_i^\dagger \hat{a}_{i+1}
+ J_i^b \hat{b}_i^\dagger \hat{b}_{i+1} +J_i^R \hat{a}_i^\dagger
\hat{b}_i +\mbox{h.c.} \right) \,.
\end{eqnarray}
Here we encounter several couplings amongst atoms of the same species
(collisional couplings $U_{aa}$, $U_{bb}$ and tunneling rates $J^a$,
$J^b$) as well as coupling of atoms of different species (collisional
coupling $U_{ab}$ and effective Raman coupling $J^R$).

We assume that initially all atoms are in the ground state
$|g_a\rangle$. This is the state which from now on will be denoted by
$|0\rangle$ whereas the second ground state $|g_b\rangle$ will serve
as the logical $|1\rangle$.

\paragraph{Single-qubit rotations}

From the above definition of the logical states it is clear that
single-qubit rotations can be performed by a Raman process. For
example, starting at $|0\rangle$ and performing half a Raman cycle
leaves us with an atom in state $|1\rangle$, whereas shorter
interactions produce superposition states between $|0\rangle$ and
$|1\rangle$. If one starts from an arbitrary superposition, the
application of a full Raman cycle generates an effective
$\hat{\sigma}_x$-rotation. In addition, a simple Rabi cycle of $2\pi$
applied to the ground state $|g_b\rangle$ is equivalent to changing
the phase of the logical $|1\rangle$ by $\pi$ and thus represents a
$\sigma_z$ operation.

\paragraph{Two-qubit gates --- controlled-phase gate and
square root of the swap operator}

\begin{figure}[ht]
\centerline{\includegraphics[width=6cm]{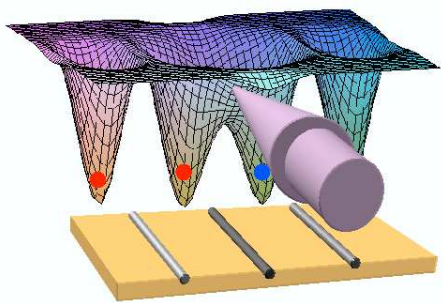}}
\caption{\label{fig:lowering} The tunneling interaction between
neighboring lattice sites can be switched on by lowering the potential
barrier between the sites.} 
\end{figure}

In order to realize two-qubit operations, we need to couple two atoms
in neighboring lattice sites to each other in a controlled way. Let us
assume that the lattice is sufficiently deep such that without
external changes tunneling between lattice sites is prohibited, hence
$J^a$ and $J^b$ can be initially neglected. A tunneling coupling between
neighboring sites can be switched on by applying an additional
standing wave perpendicular to the lattice with a waist that should
not exceed the size of two lattice sites (see
Fig.~\ref{fig:lowering}).
Depending on the circular polarization ($\sigma_+$ or $\sigma_-$) of
the additional laser beam the tunneling strength for both atomic
species can be tuned independently.

For the two-qubit operations to work, the tunneling coupling must
always be much smaller than the collisional coupling, even when the
potential barrier has been lowered. Hence, we require $U\gg J$ for all
times during the interaction process. This amounts to the fact that
the system stays in the eigenspace of the collisional terms $U_{aa}$
and $U_{bb}$, hence in a number state. This eigenspace is in fact
degenerate with respect to the occupation numbers $n_{a,b}=0$ and
$n_{a,b}=1$ since the collisional energy is obviously zero in both
cases. There is an energy gap from this degenerate subspace to the
state with two atoms of the same species per lattice site which means
that those states can be adiabatically eliminated from the
evolution.
\index{adiabatic approximation}
If, in addition, there is a large inter-species collisional
coupling $U_{ab}$, then the same energy gap persists even for states
in which two atoms of different species occupy one lattice site. Thus,
every state with more than one atom per lattice site, regardless of their
species, can be adiabatically eliminated thereby leaving us with a
degenerate eigenspace spanned by the logical states
$|0\rangle=|n_a=1,n_b=0\rangle$ and
$|1\rangle=|n_a=0,n_b=1\rangle$.
This constitutes our well-defined computational space.

\begin{figure}[ht]
\begin{minipage}{5cm}
\includegraphics[width=5cm]{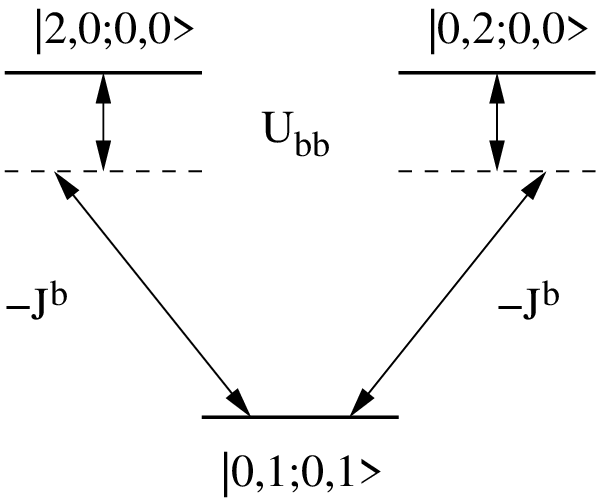}
\end{minipage}
\hfill
\begin{minipage}{5.8cm}
\includegraphics[width=5.8cm]{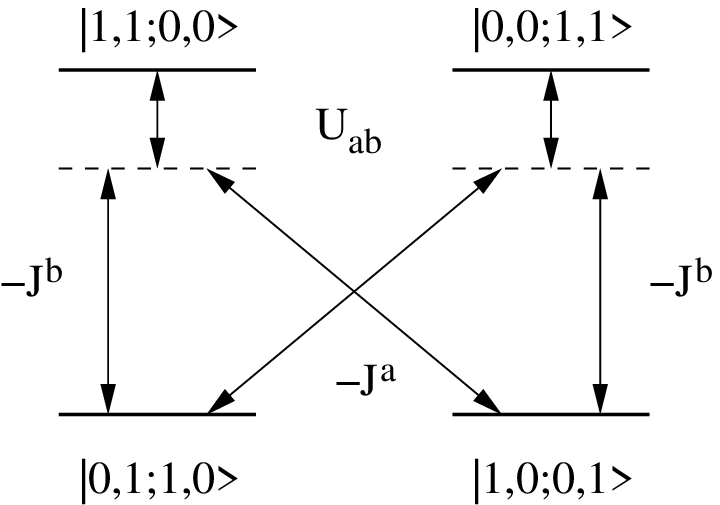}
\end{minipage}
\caption{\label{fig:levels} Effective 3-level system coupling the
state $|0,1;0,1\rangle$ to its excited states (left figure), and
effective 4-level system coupling $|0,1;1,0\rangle$ and
$|1,0;0,1\rangle$ to their respective excited states (right figure).}
\end{figure}

In what follows, we will use the notation introduced in
\cite{Pachos03} by denoting the state of the atomic population in two
lattice sites by $|n_a^1,n_b^1;n_a^2,n_b^2\rangle$. Then, for example,
a state of two atoms in their respective ground states $|g_b\rangle$
is given in this notation by $|01;01\rangle$ and represents the
logical two-qubit state $|11\rangle$. Suppose now we were to lower the
potential barrier between two neighboring sites only for the atom in
ground state $|g_b\rangle$ which can be done by choosing an
appropriate polarization of the incident laser beam. Then, according
to the Hamiltonian (\ref{eq:bosehubbard2}), this state can couple to
only two other states, $|02;00\rangle$ and $|00;02\rangle$, with two
atoms simultaneously at one lattice site. The Hamiltonian
(\ref{eq:bosehubbard2}) can be written in the basis
$\{|01;01\rangle,|02;00\rangle,|00;02\rangle\}$ as
\begin{equation}
H_{bb} = \left( \begin{array}{ccc}
0 & -J^b & -J^b \\ -J^b & U_{bb} & 0 \\ -J^b & 0 & U_{bb}
\end{array} \right) \,.
\end{equation}
This Hamiltonian effectively corresponds to a $V$-system with ground
state $|01;01\rangle$ and excited states $|02;00\rangle,|00;02\rangle$
coupled by an effective Rabi frequency $-J^b/2$ and detuned by
$U_{bb}$ (see left figure in Fig.~\ref{fig:levels}).
\index{effective V-system}
Assuming that the detuning is large, the system remains
in the ground state and acquires a phase $-2\int dt\,(J^b)^2/U_{bb}$.
However, in the same manner the logical states
$|01\rangle=|1,0;0,1\rangle$ and $|10\rangle=|0,1;1,0\rangle$ acquire
phases. This is due to their interaction with the states
$|1,1;0,0\rangle$ and $|0,0;1,1\rangle$, respectively. For example, in
the basis $\{|1,0;0,1\rangle,|1,1;0,0\rangle\}$ the Hamiltonian
(\ref{eq:bosehubbard2}) simply reads
\begin{equation}
H_{ab} = \left( \begin{array}{cc}
0 & -J^b \\ -J^b & U_{ab}
\end{array} \right)
\end{equation}
which leads to a phase shift of $\int dt\,(J^b)^2/U_{ab}$. The latter
can be compensated for by applying a single-qubit rotation on the
state $|1\rangle$ on both lattice sites. The overall effect is to
introduce a phase shift
\begin{equation}
\varphi = 2\int\limits_0^T dt\, \left( \frac{(J^b)^2}{U_{ab}}
-\frac{(J^b)^2}{U_{bb}} \right)
\end{equation}
to the logical state $|11\rangle$ while keeping all other logical
basis states unchanged. This is the controlled-phase gate which,
together with the single-qubit rotations, constitutes a universal set
of operations for quantum computing.

Similarly, we can choose to act upon the logical states $|01\rangle$
and $|10\rangle$ which contain one atom of each species in neighboring
lattice sites. In the adiabatic approximation, the Hamiltonian
connecting these two states is (in the basis
$\{|1,1;0,0\rangle,|1,0;0,1\rangle,|0,1;1,0\rangle,|0,0;1,1\rangle\}$,
see right figure in Fig.~\ref{fig:levels})
\begin{equation}
H_{ab} = \left(\begin{array}{cccc}
U_{ab} & -J^a & -J^b & 0 \\ -J^a & 0 & 0 & -J^b \\ -J^b & 0 & 0 & -J^a
\\ 0 & -J^b & -J^a & U_{ab}
\end{array}\right)
\end{equation}
with the solution that the following transformation on the logical
states $|01\rangle$ and $|10\rangle$ is achieved:
\begin{eqnarray}
|01\rangle & \mapsto & e^{-i\varphi}
\left[ \cos I |01\rangle -i\sin I |10\rangle \right] \\
|10\rangle & \mapsto & e^{-i\varphi}
\left[ -i\sin I |01\rangle +\cos |10\rangle \right]
\end{eqnarray}
with
\begin{equation}
\varphi = \int\limits_0^T dt\,\frac{(J^a)^2+(J^b)^2}{U_{ab}} \,, \quad
I = 2\int\limits_0^T dt\,\frac{J^aJ^b}{U_{ab}} \,.
\end{equation}
The effective Hamiltonian is therefore just
\begin{equation}
H_{\scriptsize\textrm{eff}} =
-I |(|10\rangle\langle 01|+|01\rangle\langle 10|)\,,
\end{equation}
which, for $I=\pi/4$ is also known also the square root of the swap
operator. Additionally, the logical states $|00\rangle$ and
$|11\rangle$ acquire phases $-2\int dt\,(J^a)^2/U_{aa}$ and
$-2\int dt\,(J^b)^2/U_{bb}$, respectively, as discussed before. Thus,
the two qubits pick up only an \textit{overall} phase if one chooses
$(J^a)^2/U_{aa}+(J^b)^2/U_{bb}=[(J^a)^2+(J^b)^2]/U_{ab}$. This overall
phase can later be reversed by applying appropriate single-qubit
rotations on both lattices sites.

These examples show that it should be straightforward in principle to
implement different types of single- and two-qubit operations with
very few laser pulses (typically just one). The drawback is that for
all above considerations the adiabaticity condition must be
fulfiled. Thus, the couplings $J^a$ and $J^b$ are assumed to be small
compared to the collisional couplings $U_{aa}$, $U_{bb}$, and $U_{ab}$
which results in unwanted long gate evolution times. For example, in
Ref.~\cite{Pachos03} it has been estimated that, for currently
measured collisional couplings of ${\cal O}$($1$kHz) \cite{Mandel03},
the gate operation time for nontrivial gates such as the
above-described controlled-phase gate with an error rate less than
$10^{-3}$ is roughly $100$ms. This time scale is far too long to
render quantum computation useful with this scheme.
A way to circumvent this problem and to drastically reduce gate
operation times is to relax the adiabaticity condition and to note
that even without adiabatic evolution there are certain instances in
which the atomic population returns completely to the logical space in
which we have started. The drawback here is that laser amplitudes and
pulse durations have to be stabilized much more precisely than in
the adiabatic regime. Despite that, it seems that there is much
potential in this and other proposals for quantum computing on an
optical lattice.

Moreover, this time scale is already of the order
of the currently possible trapping lifetime of atoms in recent
experiments (see Sec.~\ref{sec:media} for a detailed discussion 
about trapping losses). Hence, only a few gate operations can be
performed before the atoms are lost from the optical lattice.

%%%%%%%%%%%%%%%%%%%%%%%%%%%%%%%%%%%%%%%%%%%%%%%%%%%%%%%%%%%%%%%%%%%%%%
\subsection{Experimental realization with atom chips}

\begin{figure}[ht]
\centerline{\includegraphics[height=6cm]{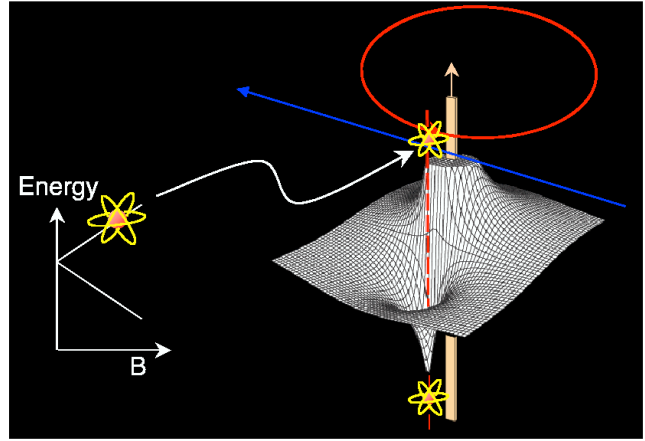}}
\caption{\label{fig:guide} Principle of the magnetic guide for atoms
(see text for details).}
\end{figure}

A particularly interesting way of implementing the above ideas of
atomic registers is by using atom chips (for reviews on this exciting
subject, see for example \cite{HindsHughes,Folman}). The basic idea
here is to trap cold atoms in one of their low-field seeking hyperfine
ground states in the combined magnetic fields of a current-carrying
wire and a constant transverse bias (see Fig.~\ref{fig:guide}).
The radial magnetic field of the wire is superimposed on a constant
homogeneous field from the side thereby creating a line of zero
magnetic field parallel to the wire. The atoms are thus trapped in a
tubular region whose distance from the wire can be tuned by adjusting
the relative strengths of the two overlapping magnetic
fields. However, as depicted in the inset of Fig.~\ref{fig:guide}, the
atoms can flip their magnetic sublevels once they reach the region of
zero magnetic field, hence yet another bias field has to be applied,
this time parallel to the wire, to prevent the atoms from doing that.

\begin{figure}[ht]
\centerline{\includegraphics[width=\textwidth]{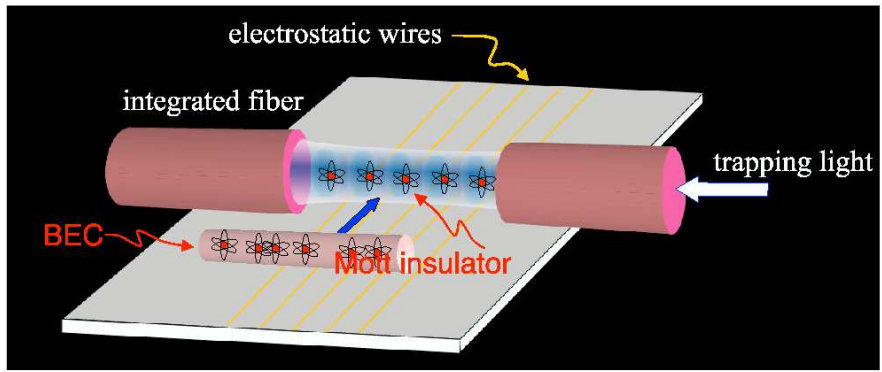}}
\caption{\label{fig:register} Proposed scheme for combining atom chips
with optical cavities for generating one-dimensional optical lattices.}
\end{figure}

Once the atoms are trapped and evaporatively cooled to form a
Bose--Einstein condensate, the atom cloud has typically a length
of 100$\mu$m and a diameter of 2$\mu$m, hence it forms a long and thin
cigar-shaped object which can be treated as quasi-1D. The idea is now
to confine the cloud between two highly reflecting mirrors that form a
microcavity in the standing wave of which the atoms experience the periodic
light potential we were envisaging earlier (Fig.\ref{fig:register}). 
The idea is then to move a string of atoms \cite{Hinds99} in a
condensate state into the cavity region to perform the transition to
the Mott-insulator. The current experimental status is that
experiments have been performed in which a cloud of trapped atoms (not
a BEC yet) has been moved in a `conveyor-belt' fashion across the
surface of a miniaturized atom chip. The next step will be to
integrate two adjacent optical fibres with polished ends onto the atom
chip to create the standing-wave potential. 

An obvious performance limitation is given by the time the atoms
actually spend in their respective trapped states above the
wire. Several noise sources, both technical and fundamental, can cause
spin flips from a trapped to an anti-trapped magnetic sublevel
that will prevent the atoms from staying indefinitely above the wire
surface. The influence of the predominant noise source --- magnetic
fluctuations caused by absorption in the current-carrying wire ---
will be investigated in more detail in Sec.~\ref{sec:media}.

%%%%%%%%%%%%%%%%%%%%%%%%%%%%%%%%%%%%%%%%%%%%%%%%%%%%%%%%%%%%%%%%%%%%%%
\section{Photonic realisation --- passive linear optics and projective
measurements}
\label{sec:photons}

In a rather different setting compared to Sec.~\ref{sec:atoms} we can
consider photons as the carriers of quantum information. Photons
constitute an alternative to atoms --- being qubits at rest --- as they
are massless particles and therefore move at the speed of light which
earned them the name `flying' qubits. They will eventually be part of
larger networks and are considered to be vital in transporting quantum
information over longer distances. Moreover, it is believed that
photons could even be used to perform quantum operations
themselves. It is therefore vital to know what kind of operations can
be done with photons and how they are implemented.

%%%%%%%%%%%%%%%%%%%%%%%%%%%%%%%%%%%%%%%%%%%%%%%%%%%%%%%%%%%%%%%%%%%%%%
\subsection{Qubit encoding and single-qubit operations}

First of all, it is necessary to define which photonic degrees of
freedom we would like to choose to encode our qubits in. One
possibility would be to encode the information in the polarization
state (horizontal or vertical) \cite{Koashi,Pittman}, another one the
superposition states of one photonic excitation in two modes
\cite{KLM}. Another, seemingly complementary encoding would use the
photon number or Fock states of a photon. All the mentioned
possibilities have  their advantages and disadvantages. For example,
single-qubit rotations are trivially implemented in the polarization
basis because they are just performed by $\lambda$/4- or
$\lambda$/2-plates. However, two-qubit operations such as the
controlled-phase gate are impossible to implement in this way using
wave plates. 

In what follows we will look at an alternative encoding in
which the qubits are defined by photon numbers. That is, the logical 0
will be the vacuum state of the electromagnetic field, and the logical
1 will be a single-photon Fock state. Then, let us consider a simple
example of a two-qubit gate, the controlled-phase gate defined by the
truth table~\ref{tab:cphase}.
That is, only the basis state containing one photon in each mode will
pick up a phase $\varphi$, all other basis states are left unchanged.

\begin{figure}[ht]
\centerline{\includegraphics[width=5cm]{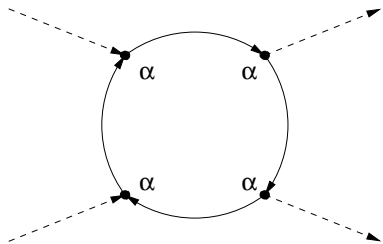}}
\caption{\label{fig:alpha4} Lowest-order Feynman diagram leading to a
photon-photon interaction.}
\end{figure}

Here we immediately encounter a problem in that, in order to realize
this quantum gate, two single photons have to interact with each other
sufficiently strongly to produce the desired phase shift. In fact,
nature is not so kind as to allow us to do this easily. Consider for a
moment standard quantum electrodynamics. The lowest-order Feynman
diagram that contains a photon-photon interaction is depicted in
Fig.~\ref{fig:alpha4}.
The interaction strength is of fourth order in the fine structure
constant $\alpha\approx 1/137$ and therefore negligible. 

A seemingly promising alternative is provided by nonlinear materials
exhibiting a $\chi^{(3)}$ or Kerr nonlinearity.
\index{Kerr nonlinearity}
The constituents of
those materials respond nonlinearly to an external electromagnetic
field, and after tracing out the matter degrees of freedom leaves
behind an effective nonlinear interaction between photons. However,
these natural nonlinearities are still too small to be of any use
since only phase shifts of the order of $10^{-8}$ can be achieved.

%%%%%%%%%%%%%%%%%%%%%%%%%%%%%%%%%%%%%%%%%%%%%%%%%%%%%%%%%%%%%%%%%%%%%%
\subsection{Measurement-induced nonlinearities}

\begin{figure}[ht]
\centerline{\includegraphics[width=4cm]{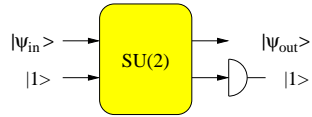}}
\caption{\label{fig:single_bs} Single beam splitter acting
conditionally as a nonlinear operator.}
\end{figure}

A possible way out of this dilemma can be found in the use of
so-called measurement-induced nonlinearities. The idea behind it is
perfectly simple. Suppose we wanted to act on a pure single-mode state
$|\psi\rangle$ with a nonlinear operator. In order to see this, we mix
our signal state with another pure single-photon state at a beam
splitter and perform a suitable projection measurement at one output
port of the beam splitter. The result will be a \textit{conditional}
nonlinear operator acting on the signal mode. For example, let us
consider the situation depicted in Fig.~\ref{fig:single_bs} in which a
single-photon Fock state acts as our auxiliary (or ancilla) state.
\index{auxiliary state}
It it well-known that a beam splitter acts as an SU(2) group element
on the photonic amplitude operators \cite{beamsplitter}. Hence, the
operator associated with the action of the beam splitter can be
written in the two-mode representation of the Pauli operators (see
Sec.~\ref{sec:intro}) as \cite{Wodkiewicz85}
\index{two-mode representation of the Pauli operators}
\begin{equation}
\label{eq:u12}
\hat{U}_{12} = T^{\hat{n}_1} e^{-R^\ast \hat{a}_2^\dagger\hat{a}_1}
e^{R\hat{a}_1^\dagger\hat{a}_2} T^{-\hat{n}_2}
\end{equation}
where $1,2$ label the modes impinging on the beam splitter. Consider
now the situation in which the auxiliary mode is prepared in a
single-mode Fock state and a single photon is detected in the same
output port. Then the resulting \textit{conditional} operator acting
on the signal mode reads
\begin{equation}
\label{eq:single_bs}
\hat{Y}_1 = \langle 1_2 | \hat{U}_{12} | 1_2 \rangle = T^{\hat{n}_1-1}
\left[ |T|^2 -\hat{n}_1 |R|^2 \right] \,.
\end{equation}
It is clear from the expression (\ref{eq:single_bs}) that this
operator represents a nonlinear evolution since it is inherently
quartic in the photonic amplitude operators
$\hat{a}_1^{(\dagger)}$. This is an \textit{effective} nonlinearity 
since the overall evolution, counting the contributions from all
possible measurement outcomes, is still perfectly \textit{linear}. 
One observes furthermore that the operator $\hat{Y}_1$ is independent of
the signal state, in fact, the nonlinearity is created by the
measurement process only. Therefore, following the notation commonly
used in quantum optics, this represents quantum-gate engineering ---
as opposed to quantum-state engineering --- or, in other words,
quantum-state engineering of arbitrary states. Note also that, on the
other hand, in general the success probability
$p=\|\hat{Y}_1|\psi\rangle\|$ does depend on the chosen signal state
$|\psi\rangle$. However, if the beam splitter parameters are chosen
such that $\hat{Y}_1^\dagger\hat{Y}_1|\psi\rangle\propto|\psi\rangle$,
i.e. the conditional operator is proportional to a unitary operator
(hence a quantum gate), then the success probability is
state-independent as well. This important fact will be used later on
to actually construct quantum gates with maximal success probability.

In order to go one step further in the development of conditional
quantum gates for photons, we have to look at larger beam splitter
networks consuming more than just one auxiliary or ancilla state. To
do so, we note that every U($N$) transformation of $N$ photonic
amplitude operators can always be realized by a triangular-shaped
network of at most $N(N-1)/2$ beam splitters and some phase shifters
\cite{Reck94}. Hence, we need to generalize our arguments to whole
networks of beam splitters. There seem to be essentially two ways of
analyzing U($N$) networks. One is to look at the unitary operator
$\hat{U}_{12\ldots N}$ acting on $N$ modes in terms of its Euler
decomposition analogous to Eq.~(\ref{eq:u12}). Although this is not
completely impossible, it is very tedious indeed. Moreover, the Euler
decomposition has only been calculated for $N=3,4$ \cite{Byrd}. An
alternative way has been developed in \cite{Scheel03} and uses
well-known techniques from bosonic operator algebras.

For this purpose, we write the input state --- still assumed to be a
single-mode state, the theory is analogous for multi-mode states --- in
a functional form as
\begin{equation}
|\psi\rangle = \hat{f}(\hat{a}_1^\dagger) |0\rangle = \sum\limits_m
\frac{c_m}{\sqrt{m!}} (\hat{a}_1^\dagger)^m |0\rangle \,,
\end{equation}
where the $c_m$ are constrained in such a way that
$\sum_m |c_m|^2=1$. Analogously, we write the auxiliary state
$|\mbox{A}\rangle$ and the state $|\mbox{P}\rangle$, respectively, in
product form as
\begin{equation}
|\mbox{A}\rangle = \prod\limits_{i=2}^N
\frac{(\hat{a}_i^\dagger)^{m_i}}{\sqrt{m_i!}} |0\rangle^{\otimes N-1}
\,,\quad
|\mbox{P}\rangle = \prod\limits_{j=2}^N
\frac{(\hat{a}_j^\dagger)^{n_j}}{\sqrt{n_j!}} |0\rangle^{\otimes N-1}
\,.
\end{equation}
Here $m_i$ is a non-negative integer that represents the number of
photons initially in mode $i$, and $n_j$ is the number of photons in
the projected mode $j$. The beam splitter network is represented by a
unitary $N\times N$-matrix $\bm{\Lambda}$, under the action of which
the amplitude operators transform as
\begin{equation}
\hat{\textbf{a}} \mapsto \bm{\Lambda}^+\hat{\textbf{a}} \,,\quad
\hat{\textbf{a}}^\dagger \mapsto \bm{\Lambda}^T
\hat{\textbf{a}}^\dagger \,.
\end{equation}
Combining all these definitions we derive the (un-normalised) output
state after mixing at the beam splitter network and projecting onto
$|\mbox{P}\rangle$ as
\begin{eqnarray}
\label{eq:mixing}
|\psi'\rangle &\propto& \langle\mbox{P}|\hat{U}_{12\ldots N}
|\mbox{A}\rangle \otimes |\psi\rangle \nonumber \\ &=&
{}^{N-1 \otimes}\langle 0| \prod\limits_{i,j=2}^N
\frac{(\hat{a}_j)^{n_j}}{\sqrt{m_i!n_j!}}
\left( \sum\limits_{k=1}^N \Lambda_{ki} \hat{a}_k^\dagger
\right)^{m_i} \hat{f}\left( \sum\limits_{l=1}^N \Lambda_{l1}
\hat{a}_l^\dagger \right) |0\rangle^{\otimes N} \,.
\end{eqnarray}
We can see immediately from Eq.~(\ref{eq:mixing}) that the effect of
the beam splitter network is to mix the photonic creation operators of
signal and auxiliary modes. At this point we use a well-known ordering
formula
\index{operator ordering formula}
\begin{equation}
\left[ \hat{a} , \hat{F}(\hat{a},\hat{a}^\dagger) \right] =
\frac{\partial}{\partial \hat{a}^\dagger}
\hat{F}(\hat{a},\hat{a}^\dagger) 
\end{equation}
to rewrite Eq.~(\ref{eq:mixing}) in the convenient form
\begin{equation}
\label{eq:output}
|\psi'\rangle \propto {}^{N-1 \otimes}\langle 0| \prod\limits_{i,j=2}^N
\frac{\left(
\frac{\partial}{\partial \hat{a}_j^\dagger}\right)^{n_j}}{\sqrt{m_i!n_j!}} 
\left( \sum\limits_{k=1}^N \Lambda_{ki} \hat{a}_k^\dagger
\right)^{m_i} \hat{f}\left( \sum\limits_{l=1}^N \Lambda_{l1}
\hat{a}_l^\dagger \right) |0\rangle^{\otimes N} \,.
\end{equation}

In the following we quote some basic results that follow immediately
from Eq.~(\ref{eq:output}). Suppose all $N-1$ auxiliary modes are
prepared in single-photon Fock states, and all $N-1$ photodetectors
find only the vacuum state at the respective output ports. Then the
output is proportional to the $N-1$-fold application of the creation
operator $(\hat{a}_1^\dagger)^{N-1}$. Similarly, if all auxiliary
modes are prepared in the vacuum state and all photodetectors find
exactly one photon each, then the output is proportional to
$\hat{a}_1^{N-1}$. These results should not be surprising if one
remembers that we are merely selecting a particular
measurement result from a passive linear operation which preserves
photon numbers. Somewhat more interesting is the situation in which
all auxiliary modes are occupied by single photons and all detectors
register a photon each. Then the associated conditional operator is a
polynomial of degree $N-1$ in the number operator $\hat{n}_1$,
$P_{N-1}(\hat{n}_1)$.  The last result is particularly important since
it allows us to act upon the (unknown) coefficients of an
$N$-dimensional signal mode \textit{independently}.

%%%%%%%%%%%%%%%%%%%%%%%%%%%%%%%%%%%%%%%%%%%%%%%%%%%%%%%%%%%%%%%%%%%%%%
\subsection{Construction of simple quantum gates}

\begin{quote}
"The permanent doesn't really interact well with linear algebra."\\
\hfill --- Mark Jerrum, private communication
\end{quote}

\begin{figure}[ht]
\centerline{\includegraphics[width=8cm]{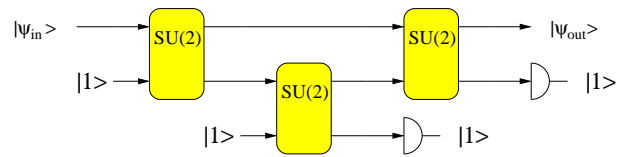}}
\caption{\label{fig:nss} SU(3) network realizing the nonlinear phase shift.}
\end{figure}

The theory outlined above presents the framework which we will use to
construct particularly simple and useful quantum gates. For
simplicity, we first restrict ourselves to single-mode gates and give
an outlook to multi-mode gates later. We also assume that we want to
operate within Fock layers, i.e. we merely change phases of some
expansion coefficients. The simplest nontrivial operations can be
generated with signal modes that contain up to two
photons.
\footnote{Note that all operations of the type
$c_0|0\rangle+c_1|1\rangle\mapsto$
$c_0|0\rangle+e^{i\varphi}c_1|1\rangle$ can be realized
deterministically, since they are merely phase shifts.} 
Let us therefore consider the nonlinear phase shift gate
$\hat{C}_\varphi$ \index{nonlinear phase shift} which is defined by
its action on a three-dimensional single-mode state as
\begin{equation}
c_0|0\rangle+c_1|1\rangle+c_2|2\rangle \stackrel{\hat{C}_\varphi}{\mapsto}
c_0|0\rangle+c_1|1\rangle+e^{i\varphi}c_2|2\rangle \,.
\end{equation}
According to what we have said earlier, in order to be able to act
upon the coefficients of this three-dimensional signal state
independently, we need a second-order polynomial in the number
operator. Hence, we immediately find a network that performs the
sought gate operation. The result is shown in Fig.~\ref{fig:nss}.
Note the triangular-shaped arrangement of the beam splitters as
predicted in \cite{Reck94}.

Let us now analyze this network along the lines described above. We
denote by $\bm{\Lambda}$ the unitary (3$\times$3)-matrix associated
with the SU(3) network. The conditional operator $\hat{Y}$ acting on
the signal mode is then
\begin{eqnarray}
\label{eq:cphase}
\lefteqn{
\hat{Y}_1 |\psi\rangle = c_0 \mbox{per}\,\bm{\Lambda}(1|1) |0\rangle
+c_1 \mbox{per}\,\bm{\Lambda} |1\rangle } \nonumber \\ &&
+c_2 (2\Lambda_{11}\mbox{per}\,\bm{\Lambda} -\Lambda_{11}^2
\mbox{per}\,\bm{\Lambda}(1|1)
+2\Lambda_{12}\Lambda_{21}\Lambda_{13}\Lambda_{31}) |2\rangle
\end{eqnarray}
where $\mbox{per}\,\bm{\Lambda}$ denotes the permanent of the matrix
$\bm{\Lambda}$ and $\mbox{per}\,\bm{\Lambda}(1|1)$ its principal 
subpermanent. \index{permanent}
In order to realize a nonlinear phase shift, we need to fulfil the
relations
\begin{equation}
\mbox{per}\,\bm{\Lambda}(1|1) = \mbox{per}\,\bm{\Lambda} \,,\quad
\mbox{per}\,\bm{\Lambda}(1|1) \left[ e^{i\varphi}
+\Lambda_{11}^2-2\Lambda_{11}  \right] =
2\Lambda_{12}\Lambda_{21}\Lambda_{13}\Lambda_{31} \,.
\end{equation}
It turns out that there are infinitely many solutions to these
equations which differ in their respective success probabilities. We
have found numerically that the maximal probability is $1/4$
\cite{Scheel03}.

\paragraph{Permanents}

The appearance of permanents in these problems is generic.
The permanent of an ($n\times n$) matrix $\bm{\Lambda}$ is a
generalized matrix function, defined as
\begin{equation}
\mbox{per}\,\bm{\Lambda}=\sum\limits_{\{\sigma_i\}\in S_n}
\prod\limits_{i=1}^n \Lambda_{i\sigma_i}
\end{equation}
where $S_n$ is the group of cyclic permutations \cite{Minc}.
Permanents naturally appear in combinatorical problems, graph theory
and related subjects. In our context they `count' the ways of
redistributing $N$  single photons through an SU($N$) network to yield
exactly $N$ single photons at the outputs, i.e.
\begin{equation}
\mbox{per}\,\bm{\Lambda} = {}^{N \otimes} \langle 1|
\hat{U}_{12\ldots N} | 1\rangle^{\otimes N} \,.
\end{equation}
As one can see from Eq.~(\ref{eq:cphase}), the unitary network has to
be adjusted such that the permanent and certain subpermanents of the
associated unitary matrix fulfil certain relations. Furthermore, the
overall success probability is just
$|\mbox{per}\,\bm{\Lambda}(1|1)|^2$. Therefore, the optimal network is
obtained by maximizing some permanent under a number of given
constraints.

This is, in fact, a nontrivial task which is partly due to the fact
that the algebraic property of a matrix being unitary does not imply
any major simplifications for computing permanents.
\footnote{This is in stark contrast to the determinant which is just
the product of the eigenvalues of a matrix.}
One of the few known fact is that permanents of unitary matrices are
bounded from above. This is a consequence of the Marcus--Newman
theorem \cite{Minc} which states that for all ($m \times n$)-matrices
$\textbf{U}$ and ($n \times m$)-matrices $\textbf{V}$ the inequality 
\index{Marcus--Newman theorem}
\begin{equation}
|\mbox{per}\,\textbf{UV}|^2 \le \mbox{per}\,\textbf{UU}^\ast
\mbox{per}\,\textbf{VV}^\ast
\end{equation}
holds which reduces, when setting $\textbf{V}=\textbf{I}$ and
regarding $\textbf{U}$ as being unitary, to
$|\mbox{per}\,\textbf{U}|\le 1$. This is, of course, what one suspects
if the permanent is supposed to be related to a success probability.

Another interesting fact to note is that computing permanents of
matrices of increasingly larger size is a computationally hard problem
in the sense that the computing time scales exponentially with the
size of the matrix. Computer scientists say this problem is
NP-complete. It seems that this computational problem is 
related to our quest to design quantum networks that would eventually
be able to solve problems in polynomial rather than exponential
time. That is, we have to solve an exponentially hard problem
\textit{first} by designing networks in order to avoid it
\textit{later} when applying them to unknown quantum states in the
course of a quantum computation. Note however, that this argument
strictly applies only if one is interested in the optimal network. For
most practical purposes it is sufficient to approximate the permanent
which can be done efficiently with a quadratic number of steps.

\paragraph{Dimension of the auxiliary state}

\begin{figure}[ht]
\centerline{\includegraphics[width=7cm]{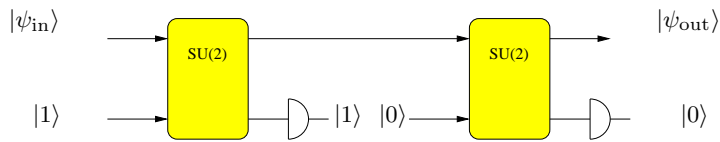}}
\begin{picture}(0,0)
\put(39,18){$|1\rangle$}
\put(153,18){$|1\rangle$}
\put(170,18){$|0\rangle$}
\put(284,18){$|0\rangle$}
\put(30,54){$|\psi_{\scriptsize\textrm{in}}\rangle$}
\put(275,54){$|\psi_{\scriptsize\textrm{out}}\rangle$}
\end{picture}
\caption{\label{fig:ralph} Nonlinear sign shift gate with two beam
splitters.}
\end{figure}

What we have said so far about the nonlinear phase shift gate regarded the
explicit construction of conditional nonlinear operators. In turns
out, however, that this way of designing quantum gates, although being
algorithmically transparent, does not yield optimal networks in terms
of resources. For example, there is an alternative network proposed in
Ref.~\cite{Ralph01} that also realizes the nonlinear phase shift but
with one fewer beam splitter (see Fig.~\ref{fig:ralph}).
As we will see later, the best networks for multi-dimensional quantum
gates in terms of their respective success probabilities are not
derived in this manner.

What seems to influence the required resources most is the
dimensionality of the used auxiliary state. Comparing
Fig.~\ref{fig:nss} and Fig.~\ref{fig:ralph} one realizes that the 
Hilbert spaces spanned by the auxiliary modes are different in both
cases. Starting from a product state of the form $|11\rangle$ as in
Fig.~\ref{fig:nss}, by beam splittings we actually span a
three-dimensional space formed of the basis states
$\{|20\rangle,|11\rangle,|02\rangle\}$. On the contrary, with an
initial product state of the form $|10\rangle$ as in
Fig.~\ref{fig:ralph} we merely span a two-dimensional space with basis
states $\{|10\rangle,|01\rangle\}$. From this observation we can
conclude that it is the dimensionality of the Hilbert space of the
auxiliary modes that determines our ability to design quantum networks
with the least possible resources. 

%%%%%%%%%%%%%%%%%%%%%%%%%%%%%%%%%%%%%%%%%%%%%%%%%%%%%%%%%%%%%%%%%%%%%%
\subsection{Multi-mode gates}

Similarly to the construction of single-mode quantum gates such as the
nonlinear phase shift, we can proceed to more complex networks by
referring to the results obtained before. The idea is rather
simple. Let us restrict our atttention again to quantum operations
within a certain Fock layer, i.e. a subspace of the total Hilbert
space of a multimode system with fixed photon number. Then we
construct an $M$-mode quantum gate in the following way:
\begin{enumerate}
\item first feed the $M$ modes into a generalized Mach--Zehnder
interferometer with $M$ input and output ports ($2M$-port for short),
\item then act upon each output mode with an appropriate single-mode
gate, 
\item and finally recombine the $M$ modes at another $2M$-port.
\end{enumerate}
Note that the first and last steps are done deterministically.

\begin{figure}[ht]
\centerline{\includegraphics[width=10cm]{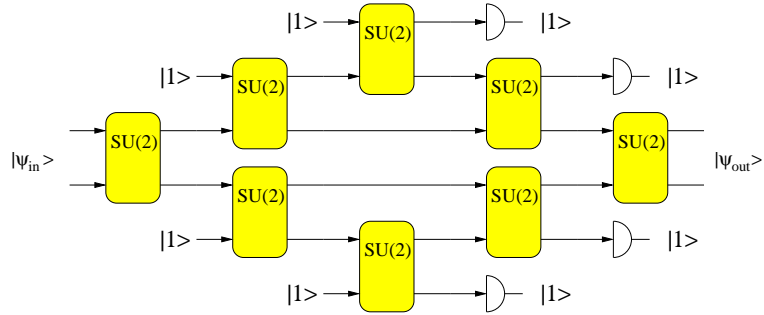}}
\caption{\label{fig:cphase} Controlled-phase gate as two nonlinear
phase shifts inside a balanced Mach--Zehnder interferometer.}
\end{figure}

Consider for example the controlled-phase gate whose truth table is
given in table~\ref{tab:cphase}. Because it represents a two-mode gate
acting in the Fock layer with total photon number equal to two, this
gate serves as the prime example for our statement. In terms of
photonic amplitude operators, the conditional operator associated with
the controlled-phase gate is given by
\begin{equation}
\hat{C}_\varphi = 1-(1-e^{i\varphi}) \hat{n}_1 \hat{n}_2 \,,
\end{equation}
where we have already written the operator in the simplest form that
matches the dimensionality of the signal-mode Hilbert spaces.
\footnote{Here we assume that the two-mode signal state is indeed only
formed by the basis states
$\{|00\rangle,|10\rangle,|01\rangle,|11\rangle\}$ without
contributions from multi-photon states.} 
If we decompose the operator $\hat{C}_\varphi$ into a product of two
unitary operators associated with the Mach--Zehnder interferometer
sandwiching some other nonlinear operator we find that these nonlinear
operator indeed forms a tensor product $\hat{N}_1\otimes\hat{N}_2$ of
two single-mode conditional operators corresponding to nonlinear phase
shift each of which has the form 
\begin{equation}
\hat{N}_i = 1-\frac{1}{2}(1-e^{i\varphi}) \hat{n}_i(\hat{n}_i-1)
\,,\quad i=1,2\,. 
\end{equation}
Thus, combining all results we have obtained so far, we end up with a
network depicted in Fig.~\ref{fig:cphase}.
With the knowledge about the maximal success probability for a
nonlinear sign flip (a nonlinear phase shift with $\varphi=\pi$) being
$1/4$, the corresponding success probability of the controlled-phase
gate will be $1/16$. 

As we have already mentioned, this is not the highest probability one
can actually achieve with linear optics and photodetection. The
`record' up until now is set by Ref.~\cite{Knill02} where a network has
been found that works with a probability of $2/27$. This network does
not have the structure we have discussed here.

%%%%%%%%%%%%%%%%%%%%%%%%%%%%%%%%%%%%%%%%%%%%%%%%%%%%%%%%%%%%%%%%%%%%%%
\subsection{Conditional dynamics and scaling of success probabilities}

In connection with the original proposal on using post-selection of
sub-dynamics it has been suggested that the use of conditional
dynamics could eventually yield success probabilities that are
arbitrary close to unity. What is meant by conditional dynamics is
that depending on a certain measurement outcome (or pattern for more
detectors) the resulting output state is either retained (in case the
measurement gave the wanted answer) or post-processed using the
knowledge about the transformation due to the detection result. We
will give a simple argument that within the framework of qubits being
encoded in photon numbers this cannot hold.

For this purpose, let us return to the single-mode gate described
earlier. Suppose therefore we had a single-mode quantum state of the
form $c_0|0\rangle+c_1|1\rangle+c_2|2\rangle$ at hand and we send it
through the arrangement of beam splitters as depicted in
Fig.~\ref{fig:ralph}. After the first beam splitter which is fed with
an auxiliary single photon the output can contain as much as three
photons in one output port. Hence, there is a certain probability that
two or three photons are found in the photodetector. Let us look at
the extreme case in which all three photons end up in the detector. By
conservation of photon numbers, the output state contains only
vacuum. But that means that we have \textit{lost all information}
about the signal state which was originally encoded in the weights
$c_i$ of the superposition. If we detected two photons instead we
would lose the information contained in $c_0$. Thus, we can conclude
that conditional dynamics breaks down if the total of number of
photons measured in all photodetectors exceeds the total number of
photons in the auxiliary state. Having said that, the obvious way of
rescuing this situation is by noting that we have taken the
measurements to be \textit{projective} measurements. If in future one
finds a way of performing non-destructive (QND) measurements, one has
immediately circumvented the information loss since all that has
happened is a unitary transformation of the signal state.

Up until now, we have to live with projective measurements and it
makes sense to ask what probabilities can be achieved in principle for
certain gates and how they scale with increasing dimensionality of the
signal state. We have already seen that the nonlinear sign flip can be
realized with $p=1/4$. In view of the algorithm to design multi-mode
gates using generalized Mach--Zehnder interferometers, we could ask
what the respective success probabilities are if we started from an
$N+1$-dimensional signal state of the form
$c_0|0\rangle+c_1|1\rangle+\ldots+c_N|N\rangle$ and we would like to
design a network that realizes a sign flip on the coefficient
$c_N$. Then in turns out that if one used an $N$-dimensional auxiliary
state with the lowest possible photon numbers, i.e. photon numbers in
the range $0\ldots N-1$, there are networks that achieve a the desired
transformation with a probability of exactly $1/N^2$ \cite{Norbert}.

%%%%%%%%%%%%%%%%%%%%%%%%%%%%%%%%%%%%%%%%%%%%%%%%%%%%%%%%%%%%%%%%%%%%%%
\section{Decoherence mechanisms --- QED in causal dielectric media}
\label{sec:media}

\begin{quote}
"Please mind the gap between the hope and the reality."\\ \hfill ---
Smoke \#2
\end{quote}

%%%%%%%%%%%%%%%%%%%%%%%%%%%%%%%%%%%%%%%%%%%%%%%%%%%%%%%%%%%%%%%%%%%%%%
\subsection{Decoherence mechanisms affecting atoms and photons}
\paragraph{Trapping losses near current-carrying wires}
As discussed in Sec.~\ref{sec:atoms}, the time the trapped atoms spend
above the current-carrying wire is limited by several loss and heating
mechanisms. In Ref.~\cite{Folman} one can find a table with several
loss mechanisms relevant for atom chip experiments. We summarize the
most important ones in table~\ref{tab:lossrates} and give their
approximate associated lifetimes.
\begin{table}
\centering
\caption{Loss mechanisms for trapped atoms above current-carrying wires}
\label{tab:lossrates}
\begin{tabular}{ll}
\hline\noalign{\smallskip}
mechanism & lifetime \\
\noalign{\smallskip}\hline\noalign{\smallskip}
thermally induced spin flips & $1-10$s \\
technical noise & $>10$s \\
background collisions & $>10$s\\
\noalign{\smallskip}\hline
\end{tabular}
\end{table}
Most of the loss mechanisms are in fact avoidable as they are merely
technical noise or account for experimental imperfections.
Technical noise can cause both heating and spin flips. Heating
processes are associated with low-frequency noise that causes the
atoms in the trap to be excited into higher vibrational modes. Spin
flips, on the other hand, are induced by rf noise. Suppose an atom
stayed 1s over the wire. The Rabi frequency that induces the spin
flips is therefore 0.5Hz, which in turn is approximately given by
$10^6$Hz/G. Therefore, the rf magnetic field must be less than
1$\mu$G which, for an atom being 1mm away from the wire, amounts to
current fluctuations of less than 1$\mu$A. For a current of $1-10$A
needed to produce the dc magnetic field this means that the power
supply must be unusually stable. This challenging but achievable.

\begin{figure}
\centering
\includegraphics[height=3cm]{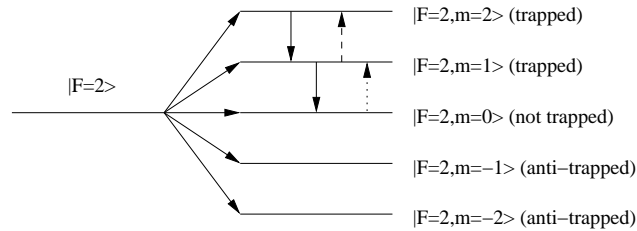}
\caption{Zeeman splitting of the $|F=2\rangle$ state of ${}^{87}$Rb
and some of the possible spin-flip transitions.}
\label{fig:sublevels}
\end{figure}

However, some loss mechanisms are inherent in the way the atoms
are trapped and can hardly be avoided. Most notable of these
unavoidable losses is that caused by thermally induced spin flips
which incidentally already gives rise to the biggest loss rate in
current experiments \cite{Jones03}. In Figure~\ref{fig:sublevels} we
show the five magnetic sublevels of the $|F\!=\!2\rangle$-state of
${}^{87}$Rb. In the experiment, the atoms are optically pumped into
the $|F\!=\!2,m\!=\!2\rangle$ sublevel in which they are trapped. Spin
flips induced by fluctutating magnetic fields cause the atom to jump
into lower-lying magnetic sublevels in which they are either lost due
to gravitational forces ($|F\!=\!2,m\!=\!0\rangle$) or expelled from
the trap ($|F\!=\!2,m\!=\!-1\rangle$ and $|F\!=\!2,m\!=\!-2\rangle$).
These spin-flip transitions are caused by magnetic-field fluctuations
that are unavoidable as soon as metallic or dielectric materials are
located close to the atoms. This is quite obviously the case in atom
chips. In order to see where these fluctuations come from, we have to
look deeper into the statistical implications of macroscopic (quantum)
electrodynamics.

\paragraph{Gate fidelity with imperfect linear optical elements}
In Sec.~\ref{sec:photons} on all-optical realisations of quantum
information processing with passive linear optical elements we have
seen that the fidelity of a gate operation will depend crucially on
the ability to generate single photons with very high efficiency and
the amount of losses that are introduced by imperfect passive elements
and photodetectors. It is not difficult to imagine that the purity of
the ancilla photons used to operate a quantum circuit will have
enormous influence on the way the quantum gates operates. For example,
an ancilla state consisting of a \textit{mixture} of vacuum and a
single-photon Fock state actually corresponds to realising
\textit{two different} quantum gates at once or, in other words, a 
statistical mixture of their respective operations. Another example of
a decoherence process is provided by the inherent absorption in the
passive linear optical elements used to build up the quantum
circuit. An absorbing beam splitter does \textit{not} act unitarily on
the electromagnetic field modes impinging on it, but rather as a
general CP map, i.e. as a statistical mixture of the desired and some
unwanted operations.

All the loss and decoherence mechanisms sketched above can be
described by a single and powerful theory which we will outline in
this section.

%%%%%%%%%%%%%%%%%%%%%%%%%%%%%%%%%%%%%%%%%%%%%%%%%%%%%%%%%%%%%%%%%%%%%%
\subsection{Field quantisation in causal media}

\begin{quote}
"Die ganzen Jahre bewu{\ss}ter Gr\"ubelei haben mich der Antwort der
Frage `Was sind Lichtquanten' nicht n\"aher gebracht. Heute glaubt
zwar jeder Lump, er wisse es, aber er t\"auscht sich."\\
\hfill --- Albert Einstein, from a letter to Michele Besso (1951)
\end{quote}

Quantum electrodynamics (or QED for short) is a well-established
theory, and one might wonder how it is possible to obtain
fundamentally new results apart from those that are already known. The
answer comes about when considering \textit{phenomenological}
electrodynamics and when trying to quantize it.

\paragraph{Classical electrodynamics with media}
Maxwell's equations of the electromagnetic field in free space,
i.e. without external sources, in temporal Fourier space can be cast
in the following form: 
\begin{eqnarray}
\label{eq:max1}
\bm{\nabla}\cdot \textbf{B}(\textbf{r},\omega) &=& 0 \,,\\
\label{eq:max2}
\bm{\nabla}\cdot \textbf{D}(\textbf{r},\omega) &=& 0 \,,\\
\label{eq:max3}
\bm{\nabla}\times \textbf{E}(\textbf{r},\omega) &=& i\omega
\textbf{B}(\textbf{r},\omega) \,,\\
\label{eq:max4}
\bm{\nabla}\times \textbf{H}(\textbf{r},\omega) &=& -i\omega
\textbf{D}(\textbf{r},\omega) \,.
\end{eqnarray}
These equations have to be supplemented by appropriate constitutive
relations which connect the electric field
\index{constitutive relations}
$\textbf{E}(\textbf{r},\omega)$ and the magnetic induction 
$\textbf{B}(\textbf{r},\omega)$ with the displacement field
$\textbf{D}(\textbf{r},\omega)$ and the magnetic field
$\textbf{H}(\textbf{r},\omega)$ which also carry information about the
material. Assuming a purely dielectric medium, i.e. disregarding
magnetic properties, one usually defines a polarisation field
$\textbf{P}(\textbf{r},\omega)$ via
\begin{equation}
\label{eq:constitutive}
\textbf{D}(\textbf{r},\omega) = \varepsilon_0
\textbf{E}(\textbf{r},\omega) + \textbf{P}(\textbf{r},\omega) \,.
\end{equation}
In real space and assuming locally and linearly responding media, the
polarisation can be written as a temporal convolution
\begin{equation}
\label{eq:linearresponse}
\textbf{P}(\textbf{r},t) = \varepsilon_0 \int\limits_0^\infty d\tau \,
\chi(\textbf{r},\tau) \textbf{E}(\textbf{r},t-\tau)
+\textbf{P}_N(\textbf{r},t) 
\end{equation}
with the linear susceptibility $\chi(\textbf{r},\tau)$, the Fourier
transform of which is related to the dielectric permittivity
$\varepsilon(\textbf{r},\omega)$ by
\index{complex dielectric permittivity}
\begin{equation}
\varepsilon(\textbf{r},\omega) = \chi(\textbf{r},\omega) +1\,.
\end{equation}
The permittivity is a complex function of frequency,
$\varepsilon(\textbf{r},\omega)$
$\!=\!\varepsilon_R(\textbf{r},\omega)$
$\!+\!i\varepsilon_I(\textbf{r},\omega)$. The real and imaginary
parts, which are responsible for dispersion and absorption,
respectively, are related to each other by the Kramers--Kronig
relations \index{Kramers--Kronig relations}
\begin{eqnarray}
\varepsilon_R(\textbf{r},\omega)-1 &=& \frac{1}{\pi} {\cal P} \int
d\omega' \frac{\varepsilon_I(\textbf{r},\omega')}{\omega'-\omega}
\,,\\
\varepsilon_I(\textbf{r},\omega) &=& -\frac{1}{\pi} {\cal P} \int
d\omega' \frac{\varepsilon_R(\textbf{r},\omega')-1}{\omega'-\omega}
\end{eqnarray}
with ${\cal P}$ denoting the principal value. The complex permittivity
is an analytic function in the upper complex half-plane without zeros
and satifies the relation
\begin{equation}
\varepsilon(\textbf{r},-\omega^\ast) =
\varepsilon^\ast(\textbf{r},\omega) \,.
\end{equation}
Furthermore, it approaches unity in the high-frequency limit. These
analyticity properties, which carry over to the Green function to be
defined later, will be important for proving equal-time
commutation relations (ETCR for short) between electromagnetic field
operators. This list of properties can be shown to follow from causality.

Equation~(\ref{eq:linearresponse}) contains an additional,
non-convolutive term $\textbf{P}_N(\textbf{r},t)$. In fact, the
existence of this term is crucial for the functioning of the following
quantisation procedure. It serves as a Langevin noise source which is
needed to preserve the ETCR. From statistical
physics it is known that dissipation processes (as described by the
imaginary part of the dielectric permittivity) are always accompanied
by additional fluctuations. In order to see this more clearly, assume
a damped harmonic oscillator with the solution for the expectation
values $\langle\hat{a}(t)\rangle$
$=\!\langle\hat{a}(t')\rangle e^{-\Gamma(t-t')}$.
This relation cannot be valid in operator form since with increasing
time the ETCR between $\hat{a}(t)$ and $\hat{a}^\dagger(t)$ would
decay exponentially and eventually violates Heisenberg's uncertainty
relations. Certainly, it is possible to add a Langevin force with
vanishing expectation value to the oscillator which takes care of the
ETCR.
\index{Langevin force}
The noise polarisation $\textbf{P}_N(\textbf{r},t)$ is exactly
such a Langevin force. Its strength is determined by the linear
fluctuation-dissipation theorem
\index{fluctuation-dissipation theorem} \cite{Stratonovich}. Assuming
that $\textbf{P}_N(\textbf{r},t)$ is proportional to a Gaussian random
variable $\textbf{f}(\textbf{r},\omega)$ which is always possible in
linear reponse theory, and noting that the linear
fluctuation-dissipation theorem states that the correlation function 
$\langle\textbf{P}(\textbf{r},\omega),\textbf{P}(\textbf{r}',\omega')$
is proportional to the imaginary part of the response function ---
hence the dielectric permittivity --- we immediately deduce that the
correct form of the noise polarisation must be
\cite{Dung98,Scheel98,Buch}
\begin{equation}
\textbf{P}_N(\textbf{r},t) = i\sqrt{\frac{\hbar\varepsilon_0}{\pi}
\varepsilon_I(\textbf{r},\omega)} \, \textbf{f}(\textbf{r},\omega)\,.
\end{equation}
As a matter of fact, the $\textbf{f}(\textbf{r},\omega)$ plays the
r\^{o}le of the fundamental variable in terms of which all relevant
quantities can be expressed. Consider for example the Helmholtz
equation for the electric field which is obtained by substituting
Faraday's law (\ref{eq:max3}) into Amper\`{e}'s law (\ref{eq:max4})
and using the constitutive relation (\ref{eq:constitutive}) as
\begin{equation}
\label{eq:helmholtz}
\bm{\nabla}\times\bm{\nabla}\times \textbf{E}(\textbf{r},\omega)
-\frac{\omega^2}{c^2} \varepsilon(\textbf{r},\omega)
\textbf{E}(\textbf{r},\omega) = \mu_0 \omega^2
\textbf{P}_N(\textbf{r},\omega) \,.
\end{equation}
The solution to Eq.~(\ref{eq:helmholtz}) is easily found to be
\begin{equation}
\label{eq:efeld}
\textbf{E}(\textbf{r},\omega) = \mu_0 \omega^2 \int d^3\textbf{s}\,
\bm{G}(\textbf{r},\textbf{s},\omega) \cdot \textbf{P}_N(\textbf{s},\omega)
\end{equation}
where the (tensor-valued) Green function satifies
\begin{equation}
\label{eq:dgl}
\bm{\nabla}\times\bm{\nabla}\times
\bm{G}(\textbf{r},\textbf{s},\omega) 
-\frac{\omega^2}{c^2} \varepsilon(\textbf{r},\omega)
\bm{G}(\textbf{r},\textbf{s},\omega) =
\delta(\textbf{r}-\textbf{s}) \bm{U} \,,
\end{equation}
where $\bm{U}$ denotes the unit dyad. Equation~(\ref{eq:efeld}) shows
explictly how all field quantities can eventually be expressed, via
the noise polarisation, in terms of the fundamental field
$\textbf{f}(\textbf{r},\omega)$. As mentioned earlier, the dyadic
Green function \index{dyadic Green function}
$\bm{G}(\textbf{r},\textbf{s},\omega)$, being a causal response
function itself, has all the analytic properties the dielectric
permittivity has, i.e. it is holomorphic in the upper complex
frequency half-plane and satifies the relation
\begin{equation}
\bm{G}(\textbf{r},\textbf{s},-\omega^\ast) =
\bm{G}^\ast(\textbf{r},\textbf{s},\omega) \,.
\end{equation}
Moreover, it fulfils a generalized Onsager--Machlup reciprocity
\cite{Stratonovich} relation of the form \index{reciprocity theorem}
\begin{equation}
\bm{G}(\textbf{s},\textbf{r},\omega) =
\bm{G}^T(\textbf{r},\textbf{s},\omega) \,.
\end{equation}
From the partial differential equation (\ref{eq:dgl}) it also follows
that the dyadic Green function satisfies an important integral
relation,
\begin{equation}
\int d^3\textbf{s} \,\frac{\omega^2}{c^2}
\varepsilon_I(\textbf{s},\omega) \bm{G}(\textbf{r},\textbf{s},\omega)
\cdot \bm{G}^+(\textbf{r}',\textbf{s},\omega) = \textrm{Im}
\bm{G}(\textbf{r},\textbf{r}',\omega) \,,
\end{equation}
which expresses the fluctuation-dissipation theorem.

\paragraph{Quantized electrodynamics with media}

The theory described above can now be quantized by regarding the
fundamental field $\textbf{f}(\textbf{r},\omega)$ as a bosonic vector
field $\hat{\textbf{f}}(\textbf{r},\omega)$ with the commutation rule
\begin{equation}
\left[ \hat{\textbf{f}}(\textbf{r},\omega) ,
\hat{\textbf{f}}^\dagger(\textbf{r}',\omega') \right] =
\delta(\textbf{r}-\textbf{r}') \delta(\omega-\omega') \bm{U} \,.
\end{equation}
Then, for example the Fourier component of the electric field operator
can be written, on using Eq.~(\ref{eq:efeld}), as
\begin{equation}
\hat{\textbf{E}}(\textbf{r},\omega) =
i \sqrt{\frac{\hbar}{\pi\varepsilon_0}} \frac{\omega^2}{c^2} \int
d^3\textbf{s} \,\sqrt{\varepsilon_I(\textbf{s},\omega)}
\bm{G}(\textbf{r},\textbf{s},\omega) \cdot
\hat{\textbf{f}}(\textbf{s},\omega) \,.
\end{equation}
The corresponding Schr\"odinger operator is obtained by integrating
over all frequency,
\begin{equation}
\hat{\textbf{E}}(\textbf{r}) = \int\limits_0^\infty d\omega
\,\hat{\textbf{E}}(\textbf{r},\omega) +\mbox{h.c.} \,.
\end{equation}
The so-defined operators fulfil all the requirements one imposes on a
statistical quantum field theory. Firstly, the ETCR between electric
field and magnetic induction can be shown to be
\cite{Dung98,Scheel98,Buch}
\begin{equation}
\left[ \hat{\textbf{E}}(\textbf{r}) , \hat{\textbf{B}}(\textbf{r}')
\right] = -\frac{i\hbar}{\varepsilon_0} \bm{\nabla}\times
\delta(\textbf{r}-\textbf{r}') \bm{U} \,.
\end{equation}
Secondly, the theory is --- by construction --- also consistent with the
fluctuation-dissipation theorem which we can cast in the form
\begin{equation}
\label{eq:fdt}
\langle 0 | \hat{\textbf{E}}(\textbf{r},\omega)
\hat{\textbf{E}}^\dagger(\textbf{r}',\omega') | 0 \rangle =
\frac{\hbar\omega^2}{\pi\varepsilon_0c^2}
\textrm{Im}\bm{G}(\textbf{r},\textbf{r}',\omega)
\delta(\omega-\omega') \,.
\end{equation}
Equation~(\ref{eq:fdt}) tells us that the (vacuum) fluctuations of the
electric field are given by the imaginary part of the Green function
as one would expect. Finally, the Hamiltonian describing the
electromagnetic field in the presence of absorbing matter is given by
\begin{equation}
\label{eq:freehamiltonian}
\hat{H} = \int d^3\textbf{r} \int\limits_0^\infty d\omega \,
\hbar\omega \,\hat{\textbf{f}}^\dagger(\textbf{r},\omega) \cdot
\hat{\textbf{f}}(\textbf{r},\omega) 
\end{equation}
which is diagonal in the fundamental bosonic vector field
$\hat{\textbf{f}}(\textbf{r},\omega)$. Maxwell's equations then follow
from the Heisenberg equations of motion of the above-defined
electromagnetic field operators.

Extensions of this theory to spatially anisotropic materials (for
which the dielectric permittivity is a symmetric tensor) and media
with gain can be found in \cite{Scheel98,Buch}. Magnetic materials,
including the important class of magnetodielectrics or left-handed
media, can be treated in a completely analogous way \cite{Dung03}. We
will, however, not elaborate more on this subject as it is not
needed for our present purposes. 

%%%%%%%%%%%%%%%%%%%%%%%%%%%%%%%%%%%%%%%%%%%%%%%%%%%%%%%%%%%%%%%%%%%%%%
\subsection{Thermally induced spin flips near metallic wires}

\begin{figure}[ht]
\centerline{\includegraphics[width=4.5cm]{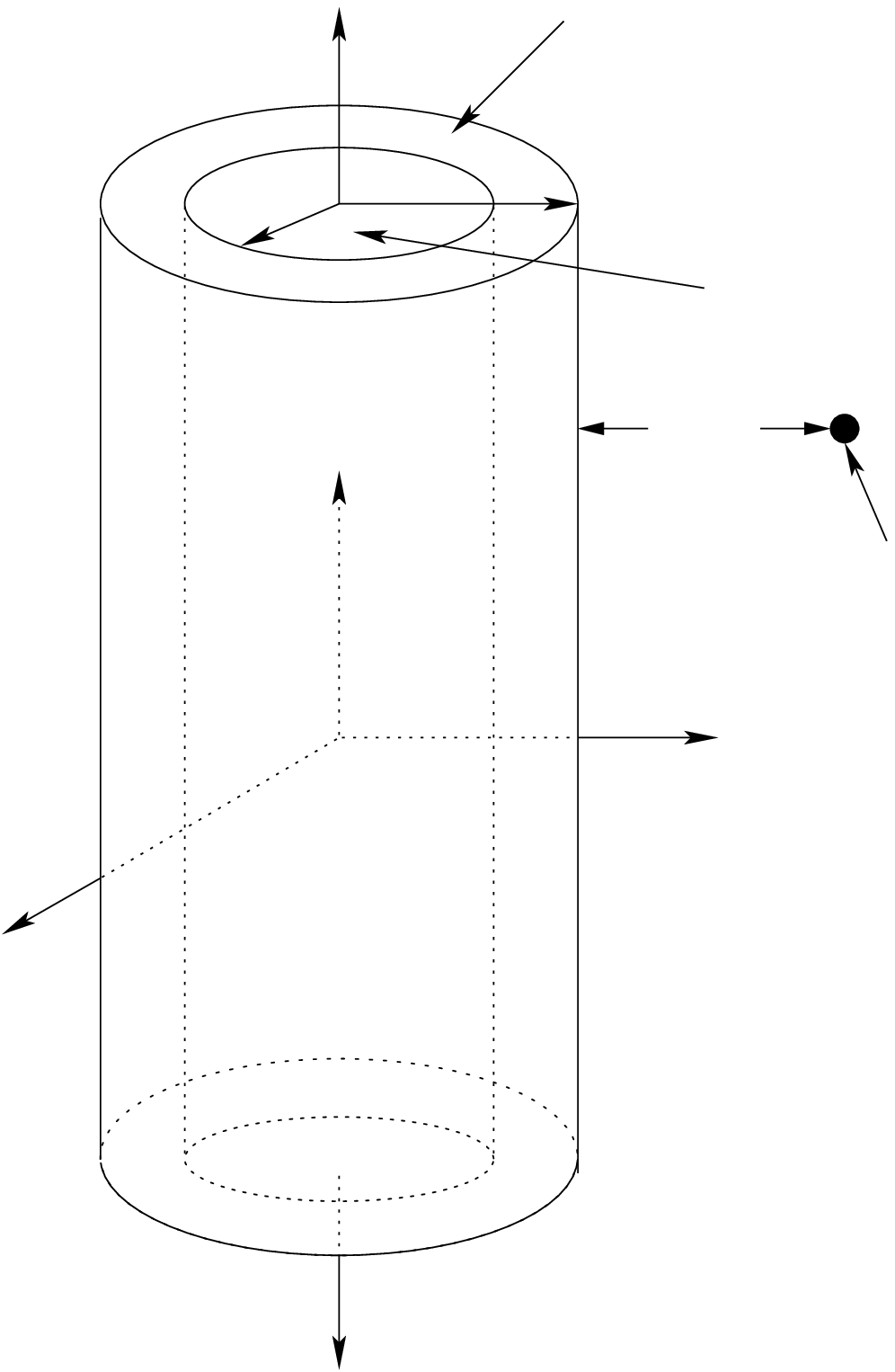}}
\unitlength=1mm
\begin{picture}(0,0)
\put(55,64){\tiny $a_2$}
\put(48,63){\tiny $a_1$}
\put(73,58){layer $1$}
\put(66,72){layer $2$}
\put(23,70){layer $3$}
\put(22,67){(vacuum)}
\put(33,25){$x$}
\put(74,35){$y$}
\put(52,51){$z$}
\put(71,51){$r$}
\put(71,43){\small two-level atom}
\end{picture}
\caption{\label{fig:wire}
A two-level atom is located at a distance $r$ from the surface of a
coated wire that runs along the $z$-direction.}
\end{figure}

The above-described theory is now perfectly suitable to describe
magnetic-field fluctuations radiated by metallic structures which
cause spin flips between trapped and anti-trapped (or non-trapped)
magnetic sublevels of atomic hyperfine ground states. The experiment
in \cite{Jones03} used a coated wire to trap ${}^{87}$Rb atoms in their
$|F\!=\!2,m\!=\!2\rangle$ magnetic sublevel. The atoms are located at a
distance $r$ from the surface of the wire with radius $a_2$ (see
Fig.~\ref{fig:wire}). 

The interaction of a neutral atom positioned at some point
$\textbf{r}_A$ with the electromagnetic field described by the
Hamiltonian (\ref{eq:freehamiltonian}) via the Zeeman interaction
$\hat{H}_Z\!=\!-\hat{\bm{\mu}}\cdot\hat{\textbf{B}}(\textbf{r}_A)$ of
the atom's magnetic moment $\hat{\bm{\mu}}$ with the magnetic field
causes the spin of the atoms to flip.
Since the atoms are in their respective (electronic) ground state, it
is safe to assume that the angular momentum is zero. Moreover, the
nuclear magnetic moment is small compared to the Bohr magneton
$\mu_B$. This leaves us with an expression for the atom's magnetic
moment associated with the transition $|i\rangle\to|f\rangle$ as
$\hat{\bm{\mu}}=\bm{\mu}|i\rangle\langle f|+\mbox{h.c.}$ where the
magnitude $\bm{\mu}$ is given by the (electronic) spin matrix element
$\bm{\mu}=2\mu_B\langle i|\hat{\textbf{S}}|f\rangle$. If we define the
spin flip operators $\hat{\sigma}=|f\rangle\langle i|$ satisfying the
angular-momentum algebra $[\hat{\sigma},\hat{\sigma}_z]=\hat{\sigma}$,
we can write the total Hamiltonian in rotating-wave approximation as
\begin{eqnarray}
\label{eq:totalhamiltonian}
\hat{H} &=& \int d^3\textbf{r} \int\limits_0^\infty d\omega \,
\hbar\omega \,\hat{\textbf{f}}^\dagger(\textbf{r},\omega) \cdot
\hat{\textbf{f}}(\textbf{r},\omega)
+\frac{1}{2}\hbar\omega_{fi} \hat{\sigma}_z \nonumber \\ && \hspace*{-7ex}
-\left[ \hat{\sigma}^\dagger \bm{\mu} \cdot \bm{\nabla}\!\times\!
\int\limits_0^\infty \!d\omega \int d^3\textbf{r} \frac{\omega}{c^2}
\,\sqrt{\frac{\hbar}{\pi\varepsilon_0}\varepsilon_I(\textbf{r},\omega)} 
\bm{G}(\textbf{r}_A,\textbf{r},\omega) \cdot
\hat{\textbf{f}}(\textbf{r},\omega) +\mbox{h.c.} \right] .
\end{eqnarray}
Using the Hamiltonian (\ref{eq:totalhamiltonian}) in the Heisenberg
equations of motion for the spin flip operators, we find after some
calculation that in the Markov approximation the spin flip rate is
given by \cite{wire}
\begin{equation}
\label{eq:barerate}
\Gamma = \frac{8\mu_B^2}{\hbar\varepsilon_0c^2}
\langle f|\hat{\textbf{S}}| i \rangle \cdot \textrm{Im} \left[
\bm{\nabla}\times\bm{\nabla}\times
\bm{G}(\textbf{r}_A,\textbf{r}_A,\omega_{fi}) \right] \cdot
\langle i|\hat{\textbf{S}}| f \rangle \,.
\end{equation}
Taking into account that the flip rate is thermally enhanced, i.e. not
only spontaneous but also stimulated (induced) flips occur, the rate
(\ref{eq:barerate}) has to be modified by a multiplicative factor
$(e^{\hbar\omega_{fi}/kT}-1)^{-1}$ which is proportional to $T$ for
high temperature. Furthermore, considering the initial state
$|i\rangle$ to be the $|F\!=\!2,m\!=\!2\rangle$ magnetic sublevel
and the final state $|f\rangle$ to be the state
$|F\!=\!2,m\!=\!1\rangle$, an additional factor of $1/4$ due to the
expansion of the $|F,m\rangle$-states in terms of the eigenstates of
electronic and nuclear spin operators is present.

\begin{figure}[ht]
\centerline{\includegraphics[width=5.5cm,angle=-90]{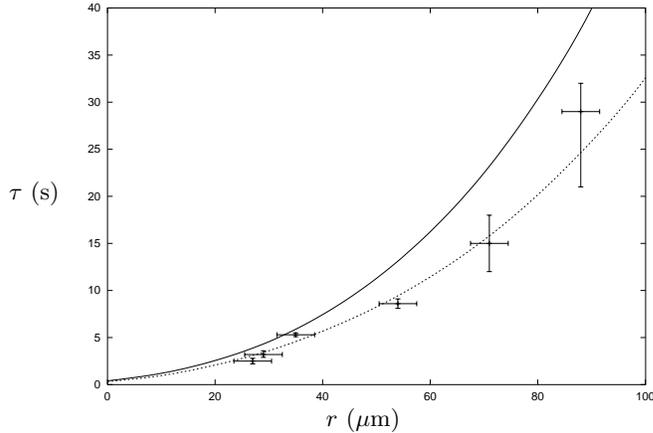}}
\begin{picture}(0,0)
\put(30,80){$\tau$ (s)}
\put(150,-6){$r$ ($\mu$m)}
\end{picture}
\caption{\label{fig:lifetime}
Lifetime $\tau$ of the trapped atom as a function of the
distance $r$ from the surface of the current-carrying wire.
The solid line corresponds to the calculated lifetime according to the
theory presented in the text. The crosses represent experimental data
points. The dotted line is the result for a slab-geometry according to
\cite{Henkel99}.} 
\end{figure}

Now we are in a position to compare with experimental
measurements. The experiment in \cite{Jones03} used a coated wire with
a copper core of 185$\mu$m radius and an aluminium coating with
55$\mu$m thickness. The transition frequency between two neighboring
magnetic sublevels was $f=\omega_{fi}/2\pi=560$kHz. The result of the
calculations with our theory and the experimental data are shown in
Fig.~\ref{fig:lifetime} where we have plotted the inverse spin flip
rate, hence the average lifetime $\tau$ of an atom in the trap.
It is clear that our theoretical result will only serve as an upper
bound on the lifetime since, as we have mentioned earlier, there are
indeed other noise sources that influence the duration of stay of the
atoms in their trap. Note, however, that the agreement is still
astonishingly good which means that our earlier claim that most other
noise source are negligible in this experiment, was indeed valid.

%%%%%%%%%%%%%%%%%%%%%%%%%%%%%%%%%%%%%%%%%%%%%%%%%%%%%%%%%%%%%%%%%%%%%%
\subsection{Imperfect passive optical elements}

Another, seemingly unrelated, application of the field quantization in
absorbing materials is found if we consider passive optical elements
such as beam splitters under realistic conditions, i.e. including
losses in the beam splitter material. As noted earlier, a lossless
beam splitter acts upon the amplitude operators of the incoming waves
as an SU(2) group element \cite{beamsplitter}. That is, two amplitude
operators $\hat{a}_1(\omega)$ and $\hat{a}_2(\omega)$ are transformed
into new amplitude operators $\hat{b}_1(\omega)$ and
$\hat{b}_2(\omega)$ as 
\begin{equation}
\left( \begin{array}{c}
\hat{b}_1(\omega) \\ \hat{b}_2(\omega) 
\end{array} \right)
= \textbf{T}(\omega)
\left( \begin{array}{c}
\hat{a}_1(\omega) \\ \hat{a}_2(\omega) 
\end{array} \right)
\end{equation}
where $\textbf{T}(\omega)$ is a unitary matrix,
i.e. $\textbf{T}(\omega)\textbf{T}^+(\omega)=\bm{I}$. For lossy beam
splitters, this is certainly not true since some of the electromagnetic
radiation impinging on the beam splitter will actually be
absorbed. This process can be accounted for by introducing an
absorption matrix $\textbf{A}(\omega)$ in order to satisfy the energy
conservation $\textbf{T}(\omega)\textbf{T}^+(\omega)+$
$\textbf{A}(\omega)\textbf{A}^+(\omega)=\bm{I}$. Both matrices
actually follow from a decomposition of the associated Green function
into contributions associated with travelling waves moving to the left
and right \cite{Gruner96} and are therefore given by the complex
refractive-index profile of the beam splitter material and geometric
parameters.

However, it turns out that, associating the material
excitations with some bosonic variables $\hat{g}_1(\omega)$ and
$\hat{g}_2(\omega)$ which are special linear combinations of the
fundamental bosonic field variables $\hat{\textbf{f}}(\textbf{r},\omega)$
restricted to one spatial dimension \cite{Gruner96}, the beam splitter
acts unitarily on the combined set of photonic and material amplitude
operators \cite{lossy}. In fact, a lossy beam splitter can be
represented as an SU(4) group element. That is, the `four-vector'
$\hat{\bm{\alpha}}(\omega)$ consisting of incoming photonic amplitude
operators and the material excitations transforms unitarily into the
`four-vector' $\hat{\bm{\beta}}(\omega)$ of outgoing photonic
amplitude operators and some new matter operators, i.e.
\begin{equation}
\hat{\bm{\beta}}(\omega) = \bm{\Lambda}(\omega)
\hat{\bm{\alpha}}(\omega) \,,\quad
\bm{\Lambda}(\omega) \bm{\Lambda}^+(\omega) = \bm{I} \,.
\end{equation}
The unitary 4$\times$4-matrix $\bm{\Lambda}(\omega)$ can be decomposed
into block form as
\begin{equation}
\bm{\Lambda}(\omega) = \left( \begin{array}{cc}
\textbf{T}(\omega) & \textbf{A}(\omega) \\
-\textbf{S}(\omega)\textbf{C}^{-1}(\omega)\textbf{T}(\omega) &
\textbf{C}(\omega)\textbf{S}^{-1}(\omega)\textbf{T}(\omega)
\end{array} \right)
\end{equation}
where
$\textbf{C}(\omega)\!=\!\sqrt{\textbf{T}(\omega)\textbf{T}^+(\omega)}$
and
$\textbf{S}(\omega)\!=\!\sqrt{\textbf{A}(\omega)\textbf{A}^+(\omega)}$
are commuting positive Hermitian 2$\times$2-matrices. The associated
unitary operator $\hat{U}$ that realizes the transformation
$\hat{\bm{\beta}}(\omega)\!=\!\bm{\Lambda}(\omega)\hat{\bm{\alpha}}(\omega)$
$\!=\!\hat{U}^\dagger\hat{\bm{\alpha}}(\omega)\hat{U}$ is then
obtained as
\begin{equation}
\hat{U} = \exp \left\{ -i\int\limits_0^\infty d\omega \left[
\hat{\bm{\alpha}}^\dagger(\omega) \right]^T  \bm{\Phi}(\omega)
\hat{\bm{\alpha}}(\omega) \right\}
\end{equation}
where the matrix $\bm{\Phi}(\omega)$ is defined by
$e^{-i\bm{\Phi}(\omega)}=\bm{\Lambda}(\omega)$. The operator $\hat{U}$
and the matrix $\bm{\Lambda}(\omega)$ can be used as well to transform
quantum states. For this purpose, let us assume that the density
operator of the quantum state before the transformation is an operator
functional of $\hat{\bm{\alpha}}(\omega)$,
i.e. $\hat{\varrho}\!=$
$\hat{\varrho}[\hat{\bm{\alpha}}(\omega),\hat{\bm{\alpha}}^\dagger(\omega)]$.
Then, transformation of the quantum state and tracing over the degrees
of freedom associated with the beam splitter leaves us with
\begin{equation}
\label{eq:statetrans}
\hat{\varrho}' = \textrm{Tr} \left\{ \hat{\varrho}\left[
\hat{U} \hat{\bm{\alpha}}(\omega) \hat{U}^\dagger,
\hat{U} \hat{\bm{\alpha}}^\dagger(\omega) \hat{U}^\dagger \right]
\right\} \,.
\end{equation}

In quantum information theory, decoherence processes are frequently
described by the Kraus decomposition of the quantum-state
transformation (\ref{eq:statetrans}).
\index{Kraus operator decomposition}
That is, we seek an operator decomposition of the form
\begin{equation}
\hat{\varrho}' = \sum\limits_i \hat{W}_i \hat{\varrho}
\hat{W}_i^\dagger \,,\quad \sum\limits_i \hat{W}_i^\dagger \hat{W}_i =
\hat{I} \,.
\end{equation}
For this purpose, let us assume that the beam splitter is in its
ground state, thereby neglecting thermal excitations which at room
temperature and optical frequencies is a safe assumption. Let us also
restrict ourselves to quasi-monochromatic radiation in which case we
can drop all frequency dependencies. Then, calculating the partial
trace in Eq.~(\ref{eq:statetrans}) in the coherent-state basis as
\begin{equation}
\hat{\varrho}' = \frac{1}{\pi^2} \int d^2\alpha_3 d^2\alpha_4 \,
\hat{W}_{\alpha_3,\alpha_4} \hat{\varrho}
\hat{W}_{\alpha_3,\alpha_4}^\dagger 
\end{equation}
where the continuous-index Kraus operators are defined by
\begin{equation}
\hat{W}_{\alpha_3,\alpha_4} = \langle \alpha_3,\alpha_4 | \hat{U}
|0_3,0_4 \rangle \,,
\end{equation}
we eventually find that the Kraus operators of an absorbing beam
splitter are ($\exp -i\bm{\Phi}_T=\textbf{T}$) \cite{Scheel03}
\begin{equation}
\label{eq:kraus}
\hat{W}_{\alpha_3,\alpha_4} =
\exp \left(-i[\hat{\textbf{a}}^\dagger]^T\bm{\Phi}_T\hat{\textbf{a}}
\right) 
\exp \left(-\bm{\alpha}^+\textbf{SC}^{-1}\textbf{T}\hat{\textbf{a}}
\right) 
\exp \left( -\bm{\alpha}^+\bm{\alpha}/2 \right) \,.
\end{equation}
One checks easily that these operators become unitary operators in the
limit of vanishing absorption, i.e. when $\textbf{T}$ is unitary and
therefore $\bm{\Phi}_T$ Hermitian, and $\textbf{S}$ vanishes. We also
see that the term
$\exp (-\bm{\alpha}^+\textbf{SC}^{-1}\textbf{T}\hat{\textbf{a}})$
is responsible for absorption as it is a function of the annihilation
operators only.

%Let us apply this theory to the simplest quantum gate imaginable, the
%nonlinear sign shift. As mentioned earlier, there are several ways to
%implement this particular gate with linear optical elements and
%photodetection. Here we choose the network proposed in \cite{Ralph01}
%for its particular simplicity (see Fig.~\ref{fig:ralph}).
%Assuming lossless beam splitters, their respective transmission
%coefficients must satisfy the relations
%\begin{equation}
%|T_{|1\rangle}|=0.476 \,,\quad |T_{|0\rangle}|=0.87 \,,\quad
%\arg T_{|1\rangle} = - \arg T_{|0\rangle} \,.
%\end{equation}
%The success probability of this gate reaches
%$|T_{|1\rangle}|^2\approx 0.23$.

\paragraph{Other error sources}

Apart from absorption in the dielectric material, there are several
other sources of errors in an experimental implementation. So far, we
have assumed that the auxiliary states are perfectly pure states,
especially we have implicitly assumed that we are able to produce
single photons on demand with very high efficiency. This, however, is
an oversimplification. All realistic sources produce as most a mixed
state of the form $(1-p)|0\rangle\langle 0|+p|1\rangle\langle 1|$ with
efficiency $p$. The best single-photon sources to date achieve
a single-photon efficiency of at most 80\%.
This means that the first beam splitter in Fig.~\ref{fig:ralph} does
not always act as a first-order polynomial in $\hat{n}$, but sometimes
subtracts a photon from the signal state.

Another crucial error source are the detectors themselves. In our
mathematical description we have assumed that the photodetectors are
perfect and can therefore be described by a projection operator
$|n\rangle\langle n|$. However, since they show losses themselves, we
have to  replace it by
\begin{equation}
|n\rangle\langle n| \mapsto
\hat{\Pi}(n) = \sum\limits_{k=n}^\infty {k \choose n}
\eta^n (1-\eta)^{k-n} |k\rangle\langle k|
\end{equation}
which constitutes a positive operator-valued measure (a generalization
of a projective measurement). It describes the effect of absorption in
the photodetector itself. The number $\eta$ is commonly called the
detector efficiency. Typical values for avalanche photo diodes are
$\eta\approx 0.3$.

We can measure the effect of these error sources by introducing the
gate fidelity $F$ as the overlap between the wanted result
$|\psi'\rangle$ and the achieved density matrix $\hat{\varrho}$,
i.e. $F=\langle\psi'|\hat{\varrho}|\psi'\rangle$. Since the so-defined
fidelity still depends on the chosen signal state --- because not all
signal states are effected in the same manner by the errors --- we
define an average fidelity $\bar{F}$ by averaging over all possible
signal states.

\begin{figure}[ht]
\centerline{\includegraphics[width=6cm]{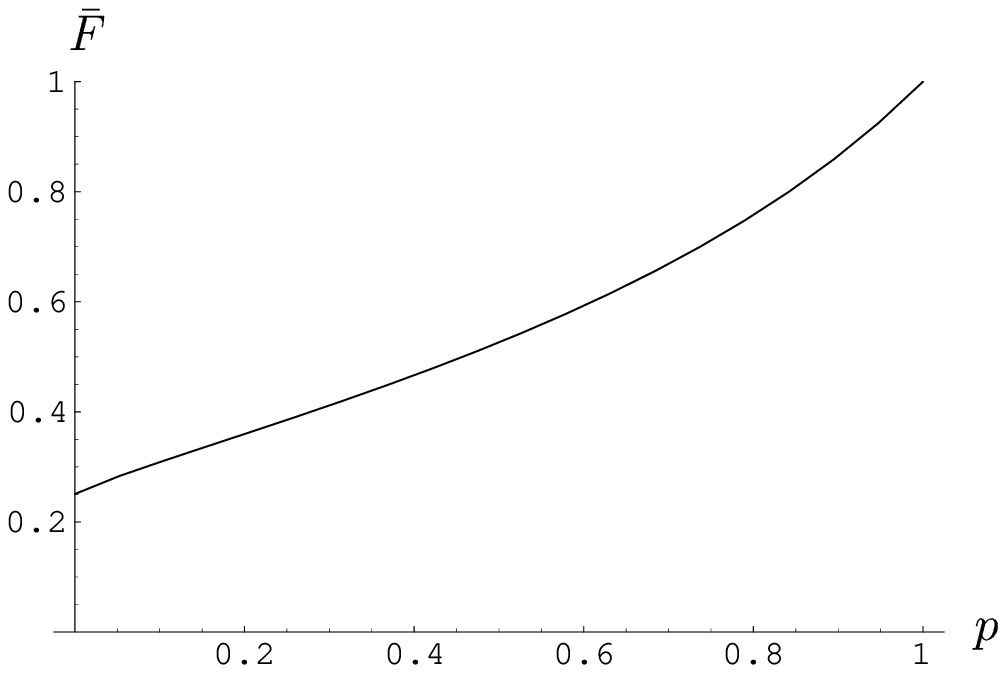}
\hfill
\includegraphics[width=6cm]{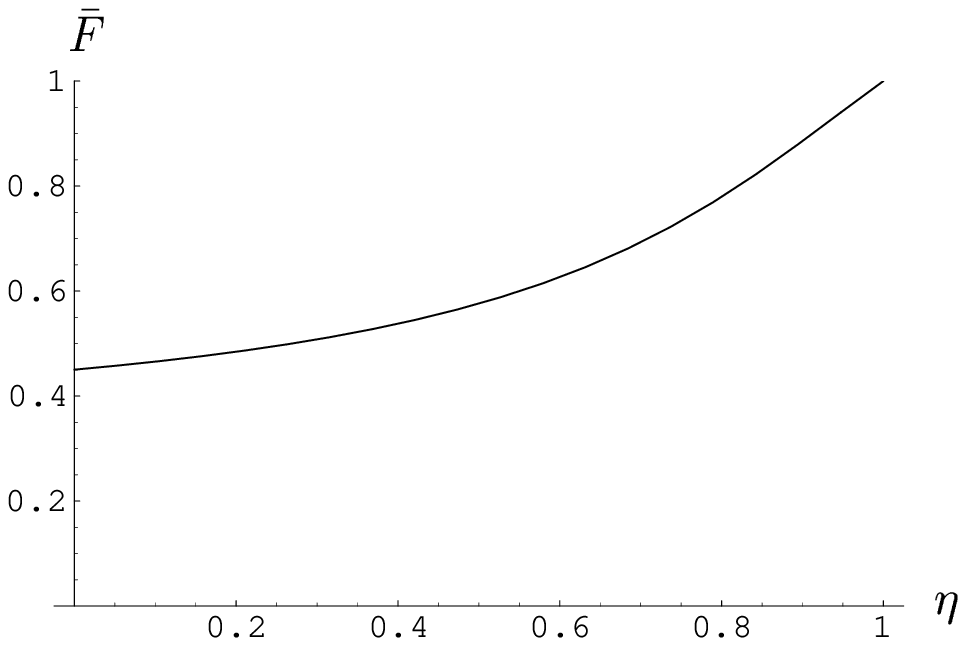}
}
\caption{\label{fig:crap} Average gate fidelity with imperfect
auxiliary photons (left figure) and imperfect detectors (right figure).}
\end{figure}

Figure~\ref{fig:crap} shows the effect of non-unit
single-photon efficiency $p$ (left figure) and detector efficiency
$\eta$ (right figure) on the average gate fidelity.
One can see that even a small inefficiency in either the detection
process or the preparation process of the auxiliary state results in a
(roughly linear) drop in the gate fidelity. In a different setting, in
which the quantum information is encoded in the superposition states
of one photonic excitation in two modes, a similar behaviour has been
observed \cite{Viv}.  This result has important
consequences for the scalability of such gates since quantum error
correction only works if the individual errors in the network
components does not exceed $\approx 10^{-3}$ (see for example
\cite{Steane03} and references cited therein). In other
words, in order to be able to correct for errors, both discussed
inefficiencies can be at most of the same order as the desired error
rate which currently seems to be experimentally impossible.

\paragraph{Acknowledgements}
We would like to thank Dominic Berry, Matthew Jones, Norbert
L\"utkenhaus, Bill Munro, Kae Nemoto, Per--Kristian Rekdal and Barry
Sanders for discussions and their input to the research reviewed here.
This work was funded in parts by the UK Engineering and Physical
Sciences Research Council (EPSRC), the Royal Society,  and the
European Commission (QUEST, QGATES, and FASTNET networks). Special
thanks goes to John, publican of the Gowlett Arms, for providing one
us with the necessary amount of Brains.

%%%%%%%%%%%%%%%%%%%%%%%%%%%%%%%%%%%%%%%%%%%%%%%%%%%%%%%%%%%%%%%%%%%%%%

%%%%%%%%%%%%%%%%%%%%%%%%%%%%%%%%%%%%%%%%%%%%%%%%%%%%%%%%%%%%%%%%%%%%%%  }

%%%%%%%%%%%%%%%%%%%%%%%%%%%%%%%%%%%%%%%%%%%%%%%%%%%%%%%%%%%%%%%%%%%%%%

\printindex
\end{document}